\documentclass{article}

\usepackage{iclr2027_conference,times}
\usepackage[T1]{fontenc}
\usepackage[utf8]{inputenc}
\usepackage{microtype}
\usepackage{amsmath,amssymb}
\usepackage{booktabs,multirow}
\usepackage{graphicx}
\usepackage{float}
\usepackage{flafter}
\usepackage{xcolor}
\usepackage{enumitem}
\usepackage{placeins}
\usepackage{xspace}
\usepackage{url}
\usepackage[hidelinks]{hyperref}
\usepackage[nameinlink,noabbrev]{cleveref}
\usepackage[font=normalsize,skip=10pt]{caption}

\definecolor{serteal}{HTML}{167D86}
\definecolor{serorange}{HTML}{D97706}
\definecolor{serblue}{HTML}{3568A8}
\definecolor{sercoral}{HTML}{C04B54}
\definecolor{sergray}{HTML}{4B5563}
\definecolor{serlight}{HTML}{F3F4F6}

\newcommand{\serbench}{\textsc{SERBench}\xspace}
\newcommand{\msscomplement}{\textsc{MSS-Complement}\xspace}
\newcommand{\ssr}{\textsc{SSR}\xspace}

\newcommand{\state}{s_t}
\newcommand{\observed}{O_t}
\newcommand{\corpus}{\mathcal{C}_r}
\newcommand{\goldgroups}{\mathcal{G}_t}

\newcommand{\Rset}{\mathcal{R}}

\newcommand{\Pset}{\mathcal{P}}
\newcommand{\best}[1]{\textbf{#1}}

\title{The Missing Complement:\\
\mbox{State-Conditioned Minimal}\\
\mbox{Sufficient Evidence for Coding Agents}}

\author{Zhexi Feng\quad Ruiyi Zhang\quad Yongbo Yang\quad Pengtao Xie\thanks{Corresponding author.}\\
Department of Electrical and Computer Engineering\\
University of California San Diego\\
La Jolla, CA 92093, USA\\
\texttt{\{zhf023,ruz048,yongboyang,p1xie\}@ucsd.edu}}

\iclrfinalcopy
\hypersetup{
  pdftitle={The Missing Complement: State-Conditioned Minimal Sufficient Evidence for Coding Agents},
  pdfauthor={Zhexi Feng, Ruiyi Zhang, Yongbo Yang, Pengtao Xie}
}


\begin{document}
\maketitle
\lhead{Preprint}

\begin{abstract}
Coding agents resolve software issues through a sequence of decisions, each requiring specific information from the repository. Standard retrieval splits the repository into small code units and scores each unit separately for relevance to a query. The top results often repeat code the agent has already read or cover only part of what it needs, wasting its limited context. To address this problem, we formulate \emph{state-conditioned minimal sufficient evidence recovery}. Given the agent's current state, we aim to retrieve the missing complement, which is a minimal sufficient set (MSS) of unread code units that together provide everything its next decision still needs. To evaluate this task, we introduce \serbench, with 500 test and 500 development states, each capturing a coding agent in the middle of solving an issue from a real repository. For each test state, expert-reviewed annotations list what the next decision still needs, and a set is complete only if it covers all of it. All methods select from the same candidate pool, which contains a complete set. We propose \msscomplement, a training-free method that chooses code units jointly. It first merges the rankings of several retrievers, and three LLM calls propose a set, extend it, and select the final 4 to 8 units within 6,144 tokens. \msscomplement returns a complete set within five units for 73.0\% of test states and within eight for 80.6\%, 11.6 and 8.2 percentage points above the strongest baseline (Qwen3 embedding with reranking). Under the same token budget, it matches this baseline's completeness with 40\% fewer tokens of code. Our task, benchmark, and method offer a new direction for accurate, context-efficient retrieval in coding agents.
\end{abstract}

\section{Introduction}
\label{sec:introduction}

\begin{figure}[!htbp]
    \centering
    \includegraphics[width=\textwidth,keepaspectratio]{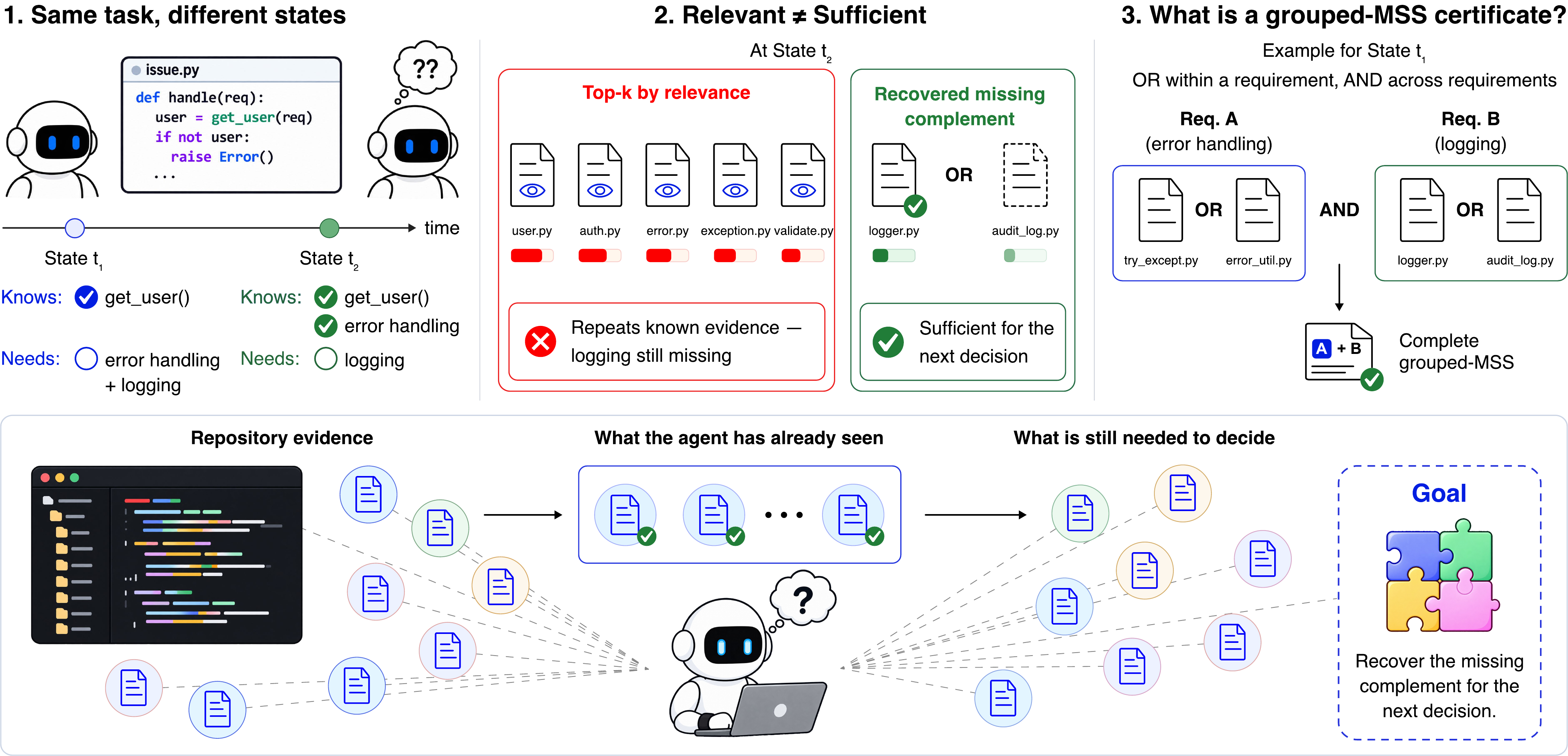}
    \caption{The code that a decision needs changes with the agent's state. Left: the agent has read \texttt{get\_user()}, and at $t_1$ its next decision needs error-handling and logging code. At $t_2$, after reading the error-handling code, it needs only logging. Middle: at $t_2$, the top results ranked by relevance repeat code the agent has already read, and logging is still missing. The missing complement is the unread code the decision still needs, here a logging file. Right: the certificate for $t_1$ lists two requirements. Either file can meet each requirement, and both requirements must be met. Bottom: the repository, the code that the agent has read, and the complement still needed for the next decision.}
    \label{fig:missing-complement}
\end{figure}

Coding agents can now resolve real software issues end to end \citep{jimenez2024swebench,yang2024sweagent,xia2024agentless}. They work step by step, searching the repository, reading code, running tests, and editing files, and each decision on what to do next depends on specific information scattered across many files. Whatever the agent reads stays in its limited context, and large language models (LLMs) often overlook relevant content as the context grows \citep{liu2024lost,hsieh2024ruler,modarressi2025nolima}. Ideally, each step should give the agent exactly the code its next decision requires. As \Cref{fig:missing-complement} shows, this code depends on the agent's state, meaning what it has done and read so far, and it often consists of complementary pieces that support the decision only together.

Standard retrievers, from embedding models and rerankers to agent-aware retrievers \citep{zhang2025qwen3embedding,shao2025reasonir,reddy2026swerank,chen2026agentir}, split the repository into small code units, such as functions or blocks of lines, and score each unit separately by its relevance to a query. Because units are scored one by one, the top results can repeat code the agent has already read. The simplest way to use the agent's state, putting it into the query, makes this worse, because the state describes what the agent has already read. In a pilot study of 21 states, such code filled 53.3\% of the top ten with this query, against 18.1\% with the issue as the query (\Cref{sec:serbench}). The results can also leave out a piece that the decision still needs, when other code looks more relevant to the query and fills the budget, as with the logging code in \Cref{fig:missing-complement}.

These failures lead to our research question: given an agent's current state, can we retrieve a small set of code that covers everything its next decision still lacks? Existing repository-retrieval benchmarks cannot answer it, because they query with the issue or workflow signals and do not record which code combinations suffice for a specific decision \citep{li2026contextbench,qin2026agentretrievalbench,zhang2026corebench,zhang2026sweexplore}. Retrieval methods for question answering, such as Evidence Tree Search and S2G-RAG, check whether evidence is sufficient, but they share little else with our task \citep{sun2025evidencetree,li2026s2grag}. They start with no evidence and serve a reader answering one fixed question, while a coding agent's need changes with every step and part of the needed code has already been read. They judge sufficiency with models trained on answer correctness or teacher labels, while an agent's decision offers no answer to learn from. They also refine the evidence over many rounds of search or retrieval, while an agent needs one compact set at each decision. Answering our question requires a benchmark that records what each decision still needs and a method that selects a sufficient set without answer supervision.

We formalize this problem as \emph{state-conditioned minimal sufficient evidence recovery} (\Cref{sec:problem}). The goal is to retrieve the missing complement for the next decision. We define it as a minimal sufficient set (MSS) of unread code units that together provide everything the decision still needs. A certificate lists the decision's requirements, and each requirement can be met by any one of several code units (right panel of \Cref{fig:missing-complement}). A set is complete only when it meets all of them.

To evaluate this task, we introduce \serbench, with 500 test states from 45 repositories and 500 development states from other repositories, all captured from coding-agent runs (\Cref{sec:serbench}). Every test state has an expert-reviewed certificate, and the main metric, Complete-MSS@$k$, is the share of states whose first $k$ returned units meet every requirement. All methods select from the same pool of 20 to 120 candidate units per state. Besides units found by retrieval, each pool includes units that the agent's later steps, the reference fix, and its tests point to, so it always contains a complete set. The main benchmark thus measures how well a method selects from given candidates, and a separate track measures how well it finds candidates in the whole repository, where no such units are added.

\serbench shows that even the strongest baseline, Qwen3 embedding with reranking \citep{zhang2025qwen3embedding}, often returns an incomplete set when a decision has several requirements. Within five units, it returns a complete set for 78.7\% of states with one requirement but for only 36.2\% of states with three or more. Because these requirements must be covered together, we propose \msscomplement, a training-free method that chooses code units jointly (\Cref{sec:method}). From a merged ranking of several retrievers, three LLM calls propose a set that covers the current need, extend it, and select the final 4 to 8 units within 6,144 tokens. Each call judges whether the chosen units together meet the need, and none sees the certificate.

On the \serbench test set, \msscomplement returns a complete set for 73.0\% of states within five units and for 80.6\% within eight, which is 11.6 and 8.2 percentage points above the strongest baseline (\Cref{sec:recovery}). On states with three or more requirements, it raises the share of complete sets from the baseline's 36.2\% to 50.3\%, a gain of 14.1 points, compared with 7.4 on single-requirement states. Joint selection thus helps most where several requirements must be covered together. The gain is 17.2 points on states whose pool contains already-read code. A control that makes the same three LLM calls but selects units by similarity reaches only 66.6\%, so the gain does not come from extra computation alone (\Cref{sec:attribution}). Under the same token budget, the method matches the baseline's completeness with 40\% fewer tokens of code, leaving more of the agent's limited context free (\Cref{sec:efficiency}).

When candidates must be found in the whole repository, all methods score lower, but \msscomplement still leads, 38.8\% to 33.8\%, so its advantage carries over while finding the needed code becomes the main bottleneck (\Cref{sec:transfer}). Whether an agent finally solves a task also depends on its model, the repository, and its earlier steps, so we test the effect of evidence at a single decision. Action52 contains 52 separate repair decisions, in each of which two LLMs read the given evidence and point to the code to change (\Cref{sec:certificate}). Removing one required piece from complete evidence and filling its place with other code lowers the precision of their choices by 11 to 12 points, so completeness itself matters. Evidence from \msscomplement raises this precision over the reranked baseline by 12.9 points for DeepSeek-v4-flash, a significant gain, and by 4.1 points for Claude Sonnet 5, so better recovery can improve the decision.

In summary, we make three contributions:
\begin{itemize}[leftmargin=*,itemsep=2pt,topsep=2pt]
\item We define state-conditioned minimal sufficient evidence recovery, the task of retrieving what an agent's next decision still lacks.
\item We introduce \serbench, with expert-reviewed certificates, a track for selection from candidate pools, and a track for retrieval from whole repositories.
\item We propose \msscomplement, a training-free method that selects code units jointly and leads all tested retrievers.
\end{itemize}
The task definition uses only a state, a pool of evidence units, and the needs of a decision, so it also applies to other agents that gather evidence step by step.

\section{Related Work}
\label{sec:related}

\paragraph{Repository context for coding agents.}
SWE-bench, SWE-agent, and Agentless evaluate or perform repository-level issue resolution, and RepoCoder studies repository-level completion by iterating retrieval and generation \citep{jimenez2024swebench,yang2024sweagent,xia2024agentless,zhang2023repocoder}. ContextBench, Agent Retrieval Bench, CORE-Bench, and SWE-Explore evaluate context and file retrieval from issue or workflow signals~\citep{li2026contextbench,qin2026agentretrievalbench,zhang2026corebench,zhang2026sweexplore}. \serbench specifies an exact intermediate decision, records the evidence established by its trajectory, and scores joint coverage of its remaining requirements.

\paragraph{Reasoning-aware and set retrieval.}
Qwen3 Embedding, ReasonIR, and SweRank study general embedding and reranking, reasoning-intensive retrieval, and software-issue localization, and AgentIR uses retrieval intent from an agent's reasoning trace \citep{zhang2025qwen3embedding,shao2025reasonir,reddy2026swerank,chen2026agentir}. Set-aware acquisition includes NEST's selection under retrieval noise, search control in RAAC and AutoSearch, and PACE's prioritization of complementary evidence \citep{verma2026nest,soudani2026raac,sun2026autosearch,cai2026pace}. In question answering, Evidence Tree Search searches sentence sets for one that lets a reader answer correctly, and S2G-RAG alternates retrieval with a judge that names missing information \citep{sun2025evidencetree,li2026s2grag}. Evidence ranking and sufficiency verification also appear in fact verification and long-video QA \citep{alt2026evidenceranking,yan2026reveal}. Our setting couples set-level acquisition to the unresolved requirements of a captured coding-agent state.

\paragraph{Agent memory and context construction.}
Generative Agents and MemGPT manage long interaction histories, while LongMemEval, LoCoMo, MemoryAgentBench, and AMA-Bench evaluate conversational, incremental, or agent-trajectory memory, and $\tau$-Knowledge ties knowledge retrieval to tool outcomes \citep{park2023generativeagents,packer2023memgpt,wu2025longmemeval,maharana2024locomo,hu2026memoryagentbench,zhao2026amabench,shi2026tauknowledge}. SCALE-QA evaluates memory in mixed-topic threads without session boundaries, where a later request depends on the right earlier episode \citep{feng2026scaleqa}. Memory methods organize software experience, revise hierarchical memories, or provide tree access, Router-Mem routes by evidence sufficiency, and MESA fuses query-specific views \citep{shen2026subtaskmemory,hu2026xmemory,liu2026semanticxpath,lin2026routermem,zhao2026mesa}. \serbench measures the combination of code units still needed at a decision, and \msscomplement supplies these units intact as the working context. Appendix~\ref{app:positioning} gives a fuller comparison.

\section{State-Conditioned Evidence Recovery}
\label{sec:problem}

A decision's working context is its visible state plus the code units acquired to support it. Within a source-token budget, the task is to acquire the missing complement, the unobserved units that together cover everything the decision still needs. We budget these units separately from the stored trajectory and from the tokens of the retrieval controller, the LLM that a retrieval method may call to select units. Let $\corpus$ be repository $r$'s frozen corpus of code units. At decision time $t$, the normalized state is $\state=(x,n_t,h_t,\observed,\psi_t)$: $x$ is the issue, $n_t$ the current information need, $h_t$ the visible-trajectory summary, $\observed\subseteq\corpus$ the code units already observed, and $\psi_t$ the hypothesis and subgoal. Let $\mathcal{V}_t\subseteq\corpus$ denote the candidate pool available at the decision, and $\mathcal{U}_t=\corpus\setminus\observed$ the unobserved evidence universe. A method returns an ordered sequence $\Rset_t=(e_1,\ldots,e_{L_t})$ from $\mathcal{V}_t$, subject to a unit limit $L_t\le K$ and the source-token budget $B$. Let $\ell(e)$ be a unit's rendered source length. The delivered context admits units in returned order while $\sum\ell(e)\le B$, and a unit that would exceed $B$ is omitted, so every scored unit stays intact. Write $\Pset_t(k)=\{e_1,\ldots,e_{\min(k,L_t)}\}$ for the first $k$ returned units.

The returned units are scored against the state's grouped certificate $\goldgroups=\{g_1,\ldots,g_{m_t}\}$, which has one group for each unresolved evidence requirement and is hidden from the method. Group $g_j=(A_j,q_j,w_j,\rho_j)$ specifies acceptable evidence IDs $A_j\subseteq\mathcal{U}_t$, their required count $q_j$, a necessity weight $w_j$, and a semantic role $\rho_j$. Each certificate specifies the support still missing at the captured decision, so its acceptable evidence is drawn from the unobserved portion of the corpus. The pool may still overlap $\observed$, and observed units keep their returned positions in every prefix, so they cost budget without covering any group. The coverage of group $g_j$ by the first $k$ returned units is
\begin{equation}
 c_j(\Rset_t,k)=\mathbb{I}\!\left[|\Pset_t(k)\cap A_j|\ge q_j\right].
\end{equation}
The threshold applies within each group and reduces to OR when $q_j=1$. All required groups are conjoined, so completion is
\begin{equation}
 C_t(k)=\prod_{j=1}^{m_t}c_j(\Rset_t,k).
 \label{eq:group-complete}
\end{equation}
A certificate also enumerates its complete alternative solutions as a family $\mathcal{B}_t$ of evidence sets, called branches. Each branch is an inclusion-minimal witness derived from the grouped certificate: every member belongs to at least one acceptable set, and every branch jointly satisfies all group thresholds. A branch is thus one minimal sufficient set (MSS) and does not widen what counts as complete. Minimality is a property of the certificate's decision-specific requirements (\Cref{sec:serbench}), so the metric asks retrieval to cover a sufficient combination within budget, and the returned set need not be provably minimal.

Over $N$ evaluation states, the primary endpoint is
\begin{equation}
 \text{Complete-MSS@}k=\frac{1}{N}\sum_{t=1}^{N}C_t(k).
 \label{eq:complete}
\end{equation}
Group recall and necessity-weighted recall measure partial coverage. Appendix~\ref{app:metrics} specifies these metrics, grouped nDCG, and prediction validation. Complete-MSS@$k$ measures recovery itself, and \Cref{sec:certificate} tests whether completion changes what the agent does next, with the state, model, tools, repository, and evaluator held fixed while only the evidence changes.

\section{SERBench}
\label{sec:serbench}

\begin{figure}[!htbp]
    \centering
    \includegraphics[width=\textwidth,keepaspectratio]{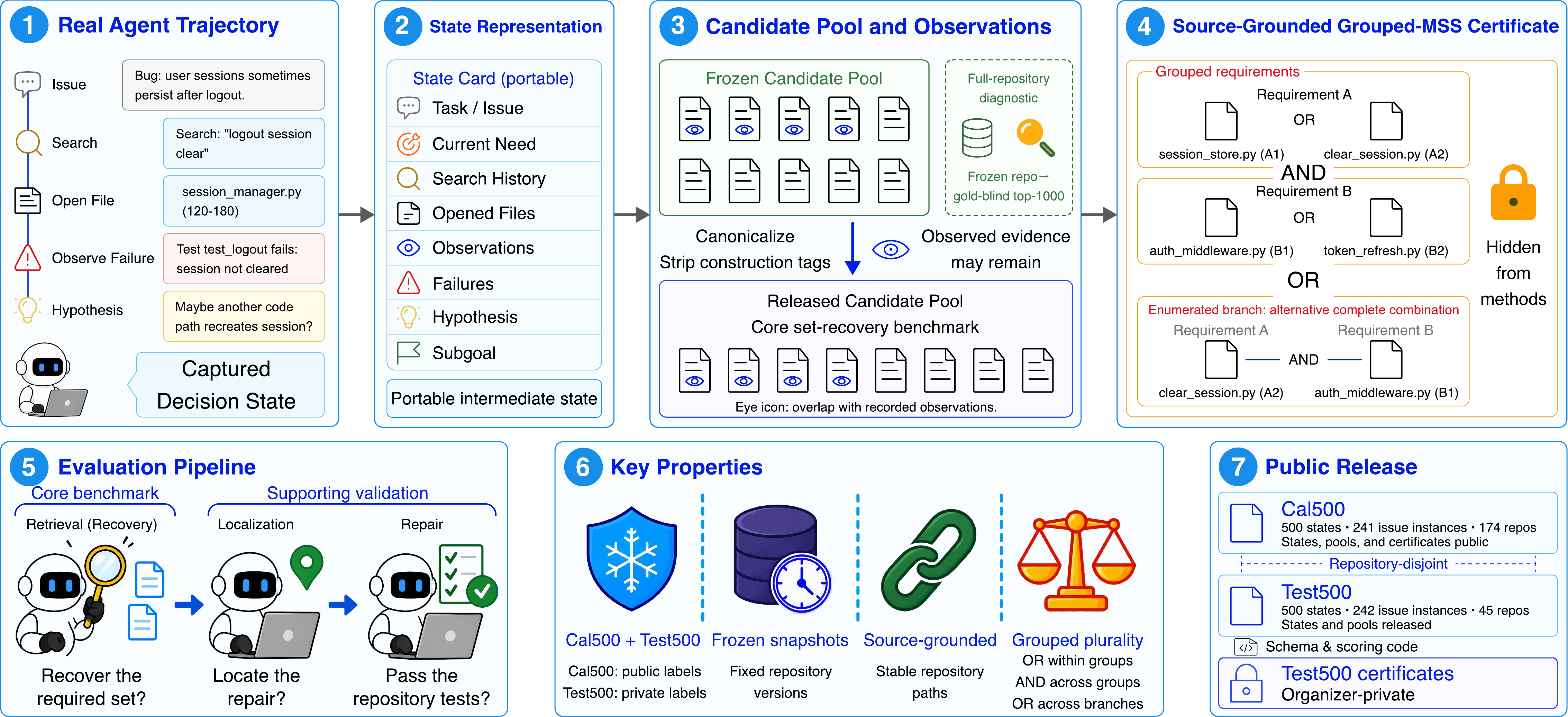}
    \caption{\serbench construction and evaluation. (1--2) A trajectory prefix becomes a portable state card. (3) Released pools drop construction tags and may contain observed code (eye icons), while a gold-blind diagnostic searches the full repository. (4) In the hidden certificate, either file meets each requirement, both requirements must be met, and the branch is one complete combination. (5) Localization and repair checks support the core set-recovery benchmark. (6--7) Only Test500 certificates stay private.}
    \label{fig:serbench-pipeline}
\end{figure}

\serbench captures a coding agent's visible state and evaluates whether retrieval recovers the code units needed for its next decision. \Cref{fig:serbench-pipeline} shows how states, certificates, and candidate pools are built and used in the recovery and downstream evaluations.

\subsection{Exact Decision States}

Test500 contains 500 states from 242 real issue instances across 45 repositories. It captures four decision boundaries: before search (200 states), after search (124), after file inspection (104), and before an intended edit (72). Each state has a reconstructable repository snapshot, issue, visible tool history, and source observations.

Trajectories come from an instrumented coding-agent harness run over repository snapshots frozen at each issue's base commit. The agent plans, searches the repository, inspects source files, and regenerates its information need, hypothesis, and subgoal after every observation. The prefix up to one of the four boundaries above is one state. Gold patches, test patches, certificates, and method outputs are never exposed while a trajectory runs. Prefixes are normalized into portable state cards carrying the current need, searches, opened files, observations, failures, hypothesis, and subgoal. Appendix~\ref{app:construction} gives the harness configuration, schema, and selection criteria.

Repository source is segmented into code units that keep their file paths and line ranges, and file reads and search results are aligned to these units to record $\observed$. Agent-authored plans remain visible context, while captured tool outputs determine which code units have already been acquired. A file path named in a plan thus enters $\observed$ only after a tool output shows its source. Test500 evaluates each frozen candidate pool as released, retaining observed units where they occur. In a 21-state pilot over the full repository index, already-read code filled 53.3\% of the top ten when the query included the agent's state, against 18.1\% with the issue as the query.

\subsection{Grouped Sufficiency Certificates}

Annotators identify the decision's unresolved requirements, ground each in code units, and specify acceptable alternatives and coverage thresholds. A requirement is retained only when its removal would weaken support for the next grounded action. Minimal sufficiency is local to the decision, allowing a repository to support several valid solutions. Groups capture jointly required facts (\Cref{eq:group-complete}), while a complete alternative branch records one sufficient combination and counts only when recovered in full (panel 4 of \Cref{fig:serbench-pipeline}). Test500 includes 124 states with multiple enumerated branches. Appendix~\ref{app:annotation} gives the annotation object, branch counts, and review protocol.

Model audits from different families first provide review signals for all certificates. Two human experts then review all 500 states independently, and a senior expert who has not seen the model decisions resolves their disagreements by checking the source. Two further experts outside that chain independently audit 80 uniformly sampled release states: both accept 77 of the 80 (96.25\%), and their judgments agree on all eight fields for 151 of 155 groups (97.42\%). Appendix~\ref{app:annotation} covers construction and adjudication, and Appendix~\ref{app:iaa} the audit.

\subsection{Candidate Pools and Release}

Released candidate pools contain 20--120 code units, mean 81.01. Besides units found by gold-blind retrieval over the whole repository, each pool includes units that the agent's later steps, the gold patch, and its test patch point to, with construction-source tags removed. Every private certificate is nonempty and grounded in its public pool, so each pool contains a complete set and the main benchmark tests selection from given candidates. The separate full-repository diagnostic tests whether a method can find the needed units from the visible state and frozen source alone. Cal500 provides 500 labeled development states from 174 repositories disjoint from Test500's. Test500 publishes state cards, bounded candidate excerpts, schemas, scoring code, and provenance fields, while certificates and evidence roles stay in a separate organizer evaluator that scores completed predictions.

\section{MSS-Complement}
\label{sec:method}

\begin{figure}[!htbp]
    \centering
    \includegraphics[width=\textwidth,keepaspectratio]{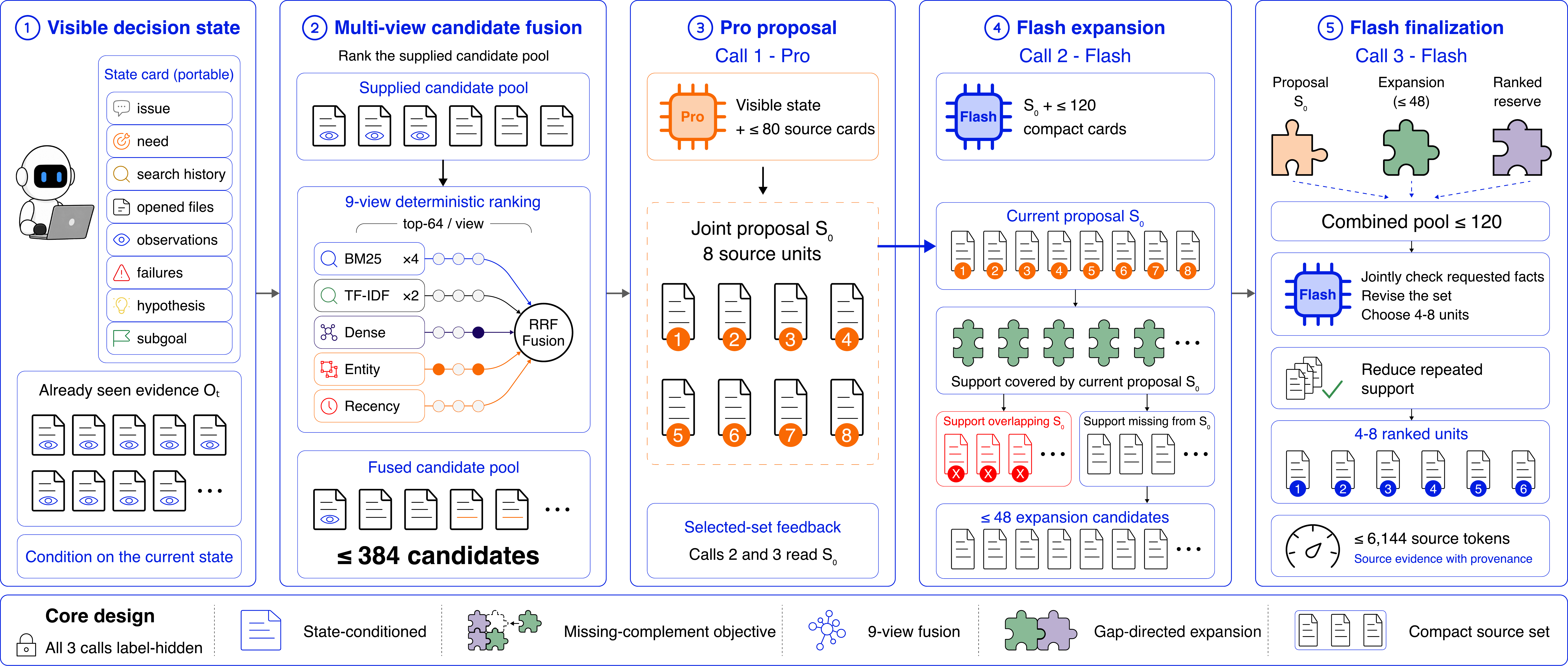}
    \caption{The \msscomplement pipeline. (1--2) Nine views rank the pool against the visible state. (3) The proposal call selects an eight-unit set $S_0$. (4) Expansion gathers up to 48 candidates for support that $S_0$ lacks. (5) Finalization returns 4 to 8 ranked units within 6,144 source tokens.}
    \label{fig:mss-complement-pipeline}
\end{figure}

\msscomplement addresses the selection step of evidence recovery. Given the visible state, a supplied candidate pool, and a token budget, it returns an ordered set of code units from the pool for the next decision. Test500 releases a pool with each state, and the full-repository diagnostic (\Cref{sec:transfer}) builds one with BM25 over the frozen repository. A merged ranking orders the pool, and three LLM calls propose a set, extend it with support the proposal lacks, and finalize it (\Cref{fig:mss-complement-pipeline}).

\subsection{Ranking the Supplied Pool}

The state card holds the issue, the current need, and the visible trajectory with its observations, hypothesis, and subgoal. Nine lexical, dense, entity, and recency views rank the pool against this state, and reciprocal-rank fusion merges their rankings into one deterministic order. The merged ranking decides which candidates each call sees and in what order, while the calls decide which units are jointly sufficient. Appendix~\ref{app:method-details} lists the views, their depths, and the pool cap.

\subsection{Three LLM Calls}

All three calls judge the chosen units as a set against the current need. The proposal call uses DeepSeek-v4-pro, abbreviated Pro. It reads the state card and up to 80 leading candidates, each as a short source card. From these it proposes an eight-unit set $S_0$ covering the entities, values, conditions, and causal links the decision needs. The expansion and finalization calls use DeepSeek-v4-flash, abbreviated Flash. Expansion is gap-directed: it reads $S_0$ and up to 120 further cards and returns up to 48 candidates for facts that $S_0$ does not yet support. Finalization reads $S_0$, these candidates, and a ranked reserve, each tagged by origin. It checks their support jointly, reduces repeated support, and returns 4 to 8 ranked units. A renderer applies the 6,144-source-token budget (\Cref{sec:problem}), so the agent receives only intact code with its provenance. \Cref{sec:attribution} measures what expansion and finalization each add.

\subsection{Operating Point and Information Access}

Controller allocation, prompts, candidate-pool and stage-input limits, and output sizes are fixed on Cal500 and transferred to the primary Test500, full-repository, AMA, prospective, and action evaluations. Ablations and controller-family comparisons vary only their declared factors. Appendix~\ref{app:method-details} records stage capacities, card limits, validation, fallback behavior, and configuration identities. Each cohort supplies its own candidate inputs, whose construction and observed-evidence handling are documented separately. The method reads only the public state and these candidates and receives no certificate, group count, gold evidence ID, construction-source tag, answer, reward, or judge output. Certificates enter only when the organizer scores completed predictions.

\section{Experiments}
\label{sec:experiments}

\Cref{sec:recovery} measures recovery of a complete evidence set under a tight budget, and \Cref{sec:efficiency} reports the context and computation this recovery uses. \Cref{sec:certificate} asks whether completing the certificate changes what an agent does next, at a repair-localization decision on Action52 and at executed repository tests on Fresh23. \Cref{sec:transfer} varies the discovery stage, task family, and controller family. \Cref{sec:attribution} attributes the gain under matched budgets.

\subsection{Common Setup}

All Test500 methods are evaluated on the same public states and frozen candidate pools. External retrievers choose their query representation on a repository-disjoint 75-state subset of Cal500, and \msscomplement uses the Cal500 configuration of \Cref{sec:method}. The supporting evaluations also use \ssr, our one-pass semantic set-selection comparator (Appendix~\ref{app:baselines}). The primary endpoint is Complete-MSS@5, with group and necessity-weighted recall, larger prefixes, and matched source-token budgets. All intervals below are 95\%. Those for Test500 and the full-repository diagnostic come from 20,000 paired resamples clustered by repository. Model endpoints, cohort-specific resampling, seeds, and trace accounting are in Appendices~\ref{app:evaluation}, \ref{app:method-details}, \ref{app:ama-accounting}, and~\ref{app:action52}.

\subsection{Recovering a Complete Evidence Set under Tight Budgets}
\label{sec:recovery}

At five units, \msscomplement reaches 73.00\% Complete-MSS, 11.60 points above Qwen3-Embedding-8B plus Qwen3-Reranker-8B, $[+6.89,+17.21]$ (\Cref{tab:main-retrieval}). Group recall improves by 9.57 points and necessity-weighted recall by 9.51, both intervals excluding zero. The same five slots thus cover more unresolved requirements and complete a combination more often. The headroom in the pools is wide: uniform selection of five units completes 5.17\% of states, and 38 points separate the strongest baseline from the oracle ceiling, of which \msscomplement closes 11.60 (Appendix~\ref{app:construction}).

At eight units the gain is 8.20 points, $[+4.91,+13.85]$. The gain is larger where several requirements must be supported together. At five units, Qwen3 + Reranker completes 78.70\% of the 216 single-group states but only 36.24\% of the 149 states needing at least three groups. \msscomplement raises these to 86.11\% and 50.34\%, a gain of 7.41 points on single-group states and 14.09 on states with at least three groups, $[+6.25,+21.67]$ (all strata in Appendix~\ref{app:group-strata}).

Because observed units cost budget without covering any group (\Cref{sec:problem}), we also split Test500 by whether the pool contains them. \msscomplement ranks first among the twelve primary methods on all five metrics in both strata, leading Qwen3 + Reranker by 8.28 points of Complete-MSS@5 in the 314 states without observed units and 17.20 in the 186 with them. The strata differ in trajectory stage, so the wider lead with observed units is descriptive only (Appendix~\ref{app:observed-overlap}).

%% FROZEN TABLE SPECIFICATION (v1.1, restored in v1.5). Full metric set is deliberate:
%% do not drop columns or rows, and do not migrate this table to the appendix.
\begin{table}[!htbp]
\centering
\caption{\serbench-Test500 at five and eight units. Every method covers all 500 states. MMR's calibrated coefficient of $1.0$ recovers the reranker ordering, so the two rows coincide (Appendix~\ref{app:baselines}). Later tables abbreviate Qwen3-Embed + Reranker-8B as Qwen3 + Reranker.}
\label{tab:main-retrieval}
\small
\setlength{\tabcolsep}{4.5pt}
\begin{tabular*}{\textwidth}{@{\extracolsep{\fill}}lrrrrr@{}}
\toprule
& \multicolumn{2}{c}{Complete} & \multicolumn{2}{c}{Group recall} & Necessity \\
\cmidrule(lr){2-3}\cmidrule(lr){4-5}\cmidrule(lr){6-6}
Method & @5 & @8 & @5 & @8 & @5 \\
\midrule
\multicolumn{6}{l}{\emph{Lexical, dense, sparse, and late interaction}} \\
BM25 & 9.00 & 13.00 & 13.12 & 18.43 & 13.16 \\
BGE-large & 31.00 & 40.80 & 40.79 & 51.83 & 40.84 \\
SPLADE++ & 23.60 & 33.40 & 31.91 & 42.29 & 32.18 \\
ColBERTv2 & 16.80 & 24.00 & 22.92 & 31.47 & 23.07 \\
\addlinespace[2pt]
\multicolumn{6}{l}{\emph{Modern general and software retrieval}} \\
ReasonIR-8B & 28.00 & 37.60 & 36.60 & 47.97 & 36.72 \\
SweRankEmbed-Large & 52.40 & 62.40 & 64.07 & 73.14 & 64.50 \\
Qwen3-Embedding-8B & 52.60 & 64.20 & 63.91 & 74.49 & 63.98 \\
Qwen3-Embed + Reranker-8B & 61.40 & 72.40 & 72.45 & 81.41 & 72.95 \\
Qwen3-Embed + Reranker-8B + MMR & 61.40 & 72.40 & 72.45 & 81.41 & 72.95 \\
\addlinespace[2pt]
\multicolumn{6}{l}{\emph{Agent-aware retrieval}} \\
AgentIR-4B & 38.60 & 47.40 & 47.51 & 57.01 & 47.71 \\
Agent-ModernColBERT & 34.00 & 43.00 & 44.00 & 54.07 & 44.45 \\
\addlinespace[2pt]
\multicolumn{6}{l}{\emph{State-conditioned set acquisition}} \\
\msscomplement & \best{73.00} & \best{80.60} & \best{82.03} & \best{87.89} & \best{82.46} \\
\bottomrule
\end{tabular*}
\end{table}

\subsection{How Much Context and Computation Does It Use?}
\label{sec:efficiency}

Every returned unit enters the agent's context, so the source tokens a method returns are a cost paid by the agent. \msscomplement returns less code than Qwen3 + Reranker while completing more states: 2,543 against 2,721 mean source tokens at five units, and 3,663 against 4,323 at eight. Under a 6,144-token budget, the reranker uses 6,100 tokens to reach 80.00\% Complete-MSS, while \msscomplement reaches 80.60\% with 3,655 tokens, 40\% fewer. When each state's token budget is matched, \msscomplement leads by 11.40 points (\Cref{tab:fair-source-budget,tab:complete-attribution-curves}).

The saving is paid for with computation outside the agent's context. On Test500 the three LLM calls use 41.50K tokens per state on average, about 1.5 US cents at off-peak November 2026 list prices (Appendix~\ref{app:method-details}), while only the final 4 to 8 units reach the agent. We consider this a favorable trade, because language models use their context less reliably as it grows longer \citep{liu2024lost,hsieh2024ruler}. \Cref{tab:external-main} gives the full token accounting on AMA-Bench.

\subsection{Does Certificate Completion Matter?}
\label{sec:certificate}
\label{sec:action52}

Test500 scores a returned set against a certificate. Whether completing it changes what an agent does next is a separate question, and two cohorts answer it. Action52 scores a repair-localization decision against validated reference repairs, and Fresh23 scores an executed repair against the repositories' own tests. Both are built by Test500's construction and screening procedure from separately frozen issue cohorts that share repositories with it (Appendix~\ref{app:construction}).

Action52 holds 52 frozen states, one from each of 52 repositories so that paired resampling has independent units. Each carries a source-grounded certificate, and 32 of the references are constructed minimal residual repairs while 20 retain the validated historical patch. Each state is paired with five evidence conditions: Qwen3-Embedding-8B, that embedding reranked, \msscomplement, Oracle-MSS, and Oracle evidence with one required group's alternatives removed. DeepSeek-v4-flash and Claude Sonnet 5 receive identical bundles and propose local code targets without repository browsing. A human reviewer checks each target against the source and labels it precise, coarse, or wrong, and fixed-budget target precision gives unused and non-precise slots zero credit (Appendix~\ref{app:action52}).

\begin{table}[!htbp]
\centering
\small
\caption{\emph{Left}, Action52 on 52 tasks: per-task evidence counts are matched within each
block under a 6,144-source-token ceiling, evidence metrics are shared by both executors, and
target precision uses a 171-slot denominator. \emph{Right}, Fresh23: the F2P column counts tasks whose fail-to-pass (F2P) tests pass, and strict resolution also needs the selected pass-to-pass tests to keep passing. Tokens is the executor's mean per-task total, prompt plus completion. Oracle rows are diagnostics built from the private certificate. Bold marks the best deployable score (Appendices~\ref{app:action52} and~\ref{app:fresh23}).}
\label{tab:action52-main}
\label{tab:fresh23-body}
\begin{minipage}[t]{0.565\textwidth}
\small
\setlength{\tabcolsep}{1.5pt}
\begin{tabular*}{\linewidth}{@{\extracolsep{\fill}}lrrrr@{}}
\toprule
& \multicolumn{2}{c}{Evidence} & \multicolumn{2}{c}{Target prec.\ (\%)} \\
\cmidrule(lr){2-3}\cmidrule(lr){4-5}
Condition & \shortstack{\strut Complete\\\strut MSS (\%)} & \shortstack{\strut Group\\\strut recall (\%)} & DeepSeek & Claude \\
\midrule
Qwen3 Embed & 21.15 & 41.60 & 48.54 & 54.39 \\
Qwen3 + Reranker & 32.69 & 52.44 & 51.46 & 54.39 \\
\msscomplement & \best{51.92} & \best{70.74} & \best{64.33} & \best{58.48} \\
\addlinespace[2pt]
Oracle $-$ 1 group & 0.00 & 48.27 & 52.63 & 55.56 \\
Oracle-MSS & 100.00 & 100.00 & 64.91 & 66.67 \\
\bottomrule
\end{tabular*}
\end{minipage}\hfill
\begin{minipage}[t]{0.415\textwidth}
\small
\setlength{\tabcolsep}{1.5pt}
\begin{tabular*}{\linewidth}{@{\extracolsep{\fill}}lrrr@{}}
\toprule
& \multicolumn{2}{c}{\shortstack{\strut Resolved\\\strut of 23}} & \\
\cmidrule(lr){2-3}
Condition & Strict & F2P & Tokens \\
\midrule
BM25 & 5 & 8 & 133,775 \\
Qwen3 + Reranker & 6 & \best{10} & 126,261 \\
\msscomplement & \best{8} & \best{10} & 118,351 \\
\addlinespace[2pt]
Oracle $-$ 1 group & 7 & 8 & 98,121 \\
Oracle-MSS & 8 & 11 & 99,655 \\
\bottomrule
\end{tabular*}
\end{minipage}
\end{table}

Removing one required group from otherwise complete evidence costs 12.28 points of precision under DeepSeek, $[+4.27,+20.34]$, and 11.11 under Claude, $[+5.20,+17.51]$ (\Cref{tab:action52-main}, left). The reduced evidence has the same count and still keeps 48.27\% group recall, so the two conditions separate on certificate completion. Among the retrievers, adding the reranker recovers much more evidence, yet neither executor's target precision separates from the embedding baseline, at $+2.92$ points for DeepSeek, $[-5.14,+11.45]$, and $0.00$ for Claude, $[-6.83,+6.63]$. Against this stronger baseline \msscomplement gains 12.87 points of precision under DeepSeek, $[+5.39,+20.36]$, and 4.09 under Claude, $[-1.89,+10.29]$. Under DeepSeek, reference-file recall improves by 4.44 points as well, $[+0.12,+10.44]$, with no loss across the 52 paired cases.

Fresh23 moves the endpoint to executed tests. It includes every qualifying task frozen after the main release, so its size is set by what was runnable. Each task runs the whole loop, in which a retrieval method supplies evidence, an executor proposes a patch, and the repository's own tests decide the outcome. Two diagnostics use the private certificate to test whether completion matters. Oracle-MSS resolves 11 of 23 tasks on the fail-to-pass (F2P) criterion, and removing one required group from its evidence cuts this to BM25's eight, losing three tasks and gaining none. \msscomplement matches Oracle-MSS on strict resolution, 8 of 23, two more than the reranked baseline, and ties that baseline on fail-to-pass (\Cref{tab:fresh23-body}, right). Every paired interval in \Cref{tab:fresh23-paired} includes zero. Retrieval is only one step of this loop, so the executor's ability, the repository's environment, and errors at earlier steps dilute its effect. This dilution and the cost of such runs, 98K to 134K executor tokens per task, are why Action52 tests evidence at a single decision, where only the evidence changes.

\subsection{Does It Transfer? Discovery, Long-Trajectory QA, and Controller Family}
\label{sec:transfer}

The full-repository diagnostic starts from frozen source using only the visible state and observed mask, over a fixed 500-state cohort that shares 434 state IDs with Test500 and carries its own certificate snapshot. Only 333 of these states have a complete set in the shared BM25 top-1,000 pool. \msscomplement reaches 38.80\% Complete-MSS@5 against 33.80\% for Qwen3 + Reranker, a gain of 5.00 points, $[+1.62,+9.44]$, from a smaller five-item source context. On the 333 states where discovery has succeeded, the gain is 7.51 points, $[+2.73,+13.88]$ (\Cref{tab:fullrepo-main}).

AMA-Bench moves the task family: 208 real agent episodes and 2,496 questions \citep{zhao2026amabench}, with the Cal500 configuration transferred without AMA-specific tuning and a single answer model and judge shared by every acquisition method. \msscomplement answers 1,399 correctly against 1,347 for AMA-Agent (\Cref{tab:external-main}), a difference of $+2.08$ points, episode-clustered $[-0.16,+4.37]$.

%% FROZEN TABLE SPECIFICATION (v1.1, restored in v1.5). Full metric set is deliberate:
%% do not drop columns or rows, and do not migrate this table to the appendix.
\begin{table}[!htbp]
\centering
\caption{AMA-208 under one answer model and judge, where token counts are means over all 2,496 questions. Answer prompt is the context visible to the answer model, answer total adds completion tokens, and online total adds controller tokens (Appendix~\ref{app:ama-accounting}).}
\label{tab:external-main}
\small
\setlength{\tabcolsep}{4pt}
\begin{tabular*}{\textwidth}{@{\extracolsep{\fill}}lrrrrrr@{}}
\toprule
& & & \multicolumn{4}{c}{Tokens per question} \\
\cmidrule(lr){4-7}
Method & Accuracy (\%) & Correct & \shortstack{Answer\\prompt} & \shortstack{Answer\\total} & Controller & \shortstack{Online\\total} \\
\midrule
BM25 & 39.58 & 988 & 2.56K & 2.67K & 0 & 2.67K \\
LongContext & 45.35 & 1,132 & 5.56K & 5.60K & 0 & 5.60K \\
Qwen3-Embedding-8B & 43.51 & 1,086 & 9.05K & 9.17K & 0 & 9.17K \\
Qwen3-Embed + Reranker-8B & 49.12 & 1,226 & 10.09K & 10.20K & 0 & 10.20K \\
\ssr & 50.04 & 1,249 & 2.34K & 2.45K & 6.45K & 8.90K \\
AMA-Agent & 53.97 & 1,347 & 13.06K & 13.95K & 0 & 13.95K \\
\msscomplement & \best{56.05} & \best{1,399} & 3.10K & 3.95K & 29.62K & 33.57K \\
\bottomrule
\end{tabular*}
\end{table}

The clearest gain is a 76.2\% smaller answer prompt alongside the higher accuracy point estimate. Constructing it consumes retrieval-controller tokens, so the online total is 33.57K against 13.95K, billed cumulatively across the retained source runs (\Cref{tab:external-main}). The result trades acquisition computation against answer-visible context. Appendix~\ref{app:ama-accounting} reports paired outcomes, deployment latency, and the full token breakdown.

Replacing the DeepSeek controller family with Claude while holding everything else fixed gives 52.27\% Complete-MSS@5 against 45.45\% on Prospective44, 44 states collected after the method was frozen, so the policy carries across families (Appendices~\ref{app:retrieval}--\ref{app:controllers} and~\ref{app:method-details}).

\subsection{Where the Gain Comes From: Attribution and Ablation}
\label{sec:attribution}

The independent-similarity control uses the same three LLM calls but selects units by similarity, without the set-sufficiency objective or selected-set feedback (the later calls reading $S_0$), and returns a fixed seven units (Appendix~\ref{app:method-details}). The full policy improves on it by 6.40 points, $[+3.42,+10.46]$ (\Cref{tab:attribution-budget}). As the formulation predicts, the two do not separate at a single returned unit, $-1.40$ points, $[-3.26,+1.05]$, and the full policy leads once the prefix has room for more than one requirement.

%% FROZEN TABLE SPECIFICATION (v1.1, restored in v1.5). Full metric set is deliberate:
%% do not drop columns or rows, and do not migrate this table to the appendix.
\begin{table}[!htbp]
\centering
\caption{Attribution and ablation on the same 500 Test500 states, from each variant's frozen predictions. The Qwen3 + Reranker row makes no controller call, the single-stage row one, and the no-expansion row two. Returned completion scores the full returned set, and context is its mean source-token count. $^\dagger$Finalization rerun in November 2026 on the frozen proposals (Appendix~\ref{app:method-details}).}
\label{tab:attribution-budget}
\small
\setlength{\tabcolsep}{2.5pt}
\begin{tabular*}{\textwidth}{@{\extracolsep{\fill}}lccrrrr@{}}
\toprule
Method & \shortstack{MSS\\objective} & \shortstack{Selection\\feedback} & Complete@5 & \shortstack{Returned\\completion} & Group@5 & Context \\
\midrule
Qwen3 + Reranker (fixed 7) & -- & -- & 61.40 & 68.80 & 72.45 & 3,781 \\
\addlinespace[2pt]
Independent similarity (fixed 7) & $\times$ & $\times$ & 66.60 & 74.40 & 76.55 & 3,595 \\
No feedback (fixed 7) & \checkmark & $\times$ & 69.80 & 78.20 & 80.07 & 3,797 \\
No feedback (adaptive 4--8) & \checkmark & $\times$ & 72.80 & 81.80 & 82.12 & 3,907 \\
\addlinespace[2pt]
Single-stage MSS proposal & \checkmark & n/a & 70.00 & 79.40 & 79.28 & 4,139 \\
No expansion (adaptive 4--8)$^\dagger$ & \checkmark & \checkmark & 72.60 & 76.40 & 81.69 & 3,085 \\
Fixed top-8 finalization & \checkmark & \checkmark & \best{73.00} & 81.40 & 82.03 & 4,113 \\
\msscomplement & \checkmark & \checkmark & \best{73.00} & 80.60 & 82.03 & 3,663 \\
\bottomrule
\end{tabular*}
\end{table}

Each step from the similarity control to the full policy changes one factor, with the candidates, call budget, and scorer fixed. The objective adds 3.20 points, $[-0.25,+8.26]$, adaptive sizing a further 3.00, $[+1.06,+6.50]$, and selected-set feedback a further 0.20, $[-4.00,+3.14]$. The objective and adaptive sizing account for 6.20 of the 6.40 points, and feedback adds no measurable gain.

To separate the two later calls, the single-stage row returns the proposal $S_0$, and the no-expansion row reruns finalization without the expansion candidates. At five units, finalization without expansion adds 2.60 points, $[0.00,+6.27]$, and expansion a further 0.40, $[-2.82,+1.93]$, for a total of 3.00, $[+0.99,+4.70]$. On the full returned set, expansion adds 4.20 points, $[+0.78,+6.30]$ (Appendix~\ref{app:method-details}). The two later calls thus matter at different budgets: finalization supplies most of the five-unit gain, and expansion stops finalization from discarding needed support on the full set. They account for about a quarter of the three calls' cost (Appendix~\ref{app:method-details}).

\FloatBarrier
\section{Conclusion}

Coding agents need the code their next decision still lacks, and this need changes with the agent's state. We formalized it as state-conditioned minimal sufficient evidence recovery, built \serbench with expert-reviewed certificates, and proposed \msscomplement, a training-free method that jointly selects code units from a supplied pool. Within five units it returns a complete set for 73.0\% of test states, 11.6 points above the strongest baseline. A control that makes the same three LLM calls but selects by similarity reaches 66.6\%, so the overall set-selection policy adds 6.4 points. With candidates from the whole repository, the method still leads, but finding them becomes the main bottleneck. Complete evidence matters downstream: removing one required piece lowers repair-localization precision under both executors, and \msscomplement raises it significantly under DeepSeek-v4-flash.

\section*{Ethics Statement}
External experts who annotated, adjudicated, or audited evaluation labels were paid for their work and agreed to the use of their judgments in this research. We assessed the need for ethics review under our institution's criteria: the experts judged public source code and recorded agent trajectories rather than providing information about themselves, so the annotation does not constitute human subjects research and institutional review was not required. Release sanitization is described in Appendix~\ref{app:artifacts}.

\section*{Reproducibility Statement}
The \serbench benchmark repository is maintained at \url{https://github.com/LordTARN1SHED/SERBench}. The supplementary archive prepared for this work provides repository-disjoint Cal500 and Test500, the scorer, method code, prompts, configurations, and aggregate predictions; Test500 certificates stay in a separate organizer evaluator that is not part of that archive. The Fresh23 end-to-end run records are included in the same archive as a separate frozen package described in Appendix~\ref{app:artifacts}. Appendices~\ref{app:construction}--\ref{app:scope} describe construction, annotation, evaluation, reproduction, and release scope. The accompanying ledger maps reported results to their configurations and source artifacts. It also includes state-level observed-overlap assignments and the script that reproduces the post-hoc subgroup summaries from frozen per-state scores.

\section*{AI Use Statement}
The authors made all research decisions. Generative AI tools assisted in implementing methods, baselines, and experiment orchestration; in constructing and formatting state cards and certificates, with human expert review as described in Appendix~\ref{app:annotation}; in giving feedback on experimental design and in analyzing and interpreting results; in translating and polishing text drafted by the authors; and in finding and checking related work and references. Models used within the research itself, including the cross-family certificate audit, the studied methods, and the models used in evaluation, are specified in Appendices~\ref{app:annotation}, \ref{app:evaluation}, and~\ref{app:method-details}. The authors reviewed all AI-assisted work and take full responsibility for this submission.

\bibliographystyle{iclr2027_conference}
\bibliography{references}

\clearpage
\setlength{\textfloatsep}{20pt plus 2pt minus 4pt}
\setlength{\intextsep}{12pt plus 2pt minus 2pt}
\setlength{\floatsep}{12pt plus 2pt minus 2pt}
\appendix
\raggedbottom
Appendices~\ref{app:positioning} and~\ref{app:construction} position \serbench and document how it was built; Appendix~\ref{app:annotation} covers certificate annotation and audit. Appendix~\ref{app:metrics} gives the evaluation protocol, the baselines, and the extended Test500 analyses; Appendix~\ref{app:method-details} specifies \msscomplement and its attribution. Appendix~\ref{app:retrieval} onward reports the cohorts outside the released pools, and Appendices~\ref{app:artifacts} and~\ref{app:scope} the artifact and its scope.

\section{Benchmark Positioning and Specification}
\label{app:positioning}

SERBench lies at the intersection of repository-context evaluation, process-grounded retrieval, and agent-memory diagnosis. ContextBench, Agent Retrieval Bench, CORE-Bench, and SWE-Explore examine context retrieval, file retrieval, and repository exploration from issue or workflow signals \citep{li2026contextbench,qin2026agentretrievalbench,zhang2026corebench,zhang2026sweexplore}. \Cref{tab:positioning} compares the properties that define its evaluation target: an exact intermediate decision as the query, already observed evidence recorded as IDs in the released candidate pool, a grouped sufficiency target, private evaluation certificates, and a paired test of evidence use.

\begin{table}[ht]
\centering
\caption{Positioning against closely related evaluation settings. The State column records the trajectory granularity used as the retrieval condition. The Observed IDs column marks whether each decision state stores the evidence its trajectory has already observed as IDs in the released candidate pool; \serbench additionally audits that overlap. ContextBench instead traces the code an evaluated agent inspects, and Agent Retrieval Bench records query-supplied files as given context. Grouped MSS requires alternatives within a requirement and conjunction across requirements.}
\label{tab:positioning}
\small
\setlength{\tabcolsep}{2.2pt}
\begin{tabular*}{\textwidth}{@{\extracolsep{\fill}}lcccccc@{}}
\toprule
Setting & Code & State & \shortstack{Observed\\IDs} & Grouped MSS & Private & Action test \\
\midrule
ContextBench~\citep{li2026contextbench} & \checkmark & process & -- & -- & -- & analysis \\
Agent Retrieval Bench~\citep{qin2026agentretrievalbench} & \checkmark & workflow & -- & -- & -- & pilot \\
MemoryAgentBench~\citep{hu2026memoryagentbench} & -- & interaction & -- & -- & -- & -- \\
AgenticRAGTracer~\citep{you2026agenticragtracer} & -- & hop trace & -- & -- & -- & diagnostic \\
\textbf{SERBench} & \checkmark & exact & \checkmark & \checkmark & \checkmark & paired \\
\bottomrule
\end{tabular*}
\end{table}

SERBench builds on established retrieval primitives (BM25, dense dual encoders, late interaction, reciprocal-rank fusion, and maximal marginal relevance) and on context engineering practice \citep{robertson2009bm25,karpukhin2020dpr,khattab2020colbert,cormack2009rrf,carbonell1998mmr,rajasekaran2025context}. Long-context evaluations document position, scale, and lexical-overlap limits \citep{liu2024lost,hsieh2024ruler,modarressi2025nolima}; SERBench instantiates these concerns at a captured software-agent decision and scores a compact source-grounded set against its unresolved requirements.

\section{SERBench Construction and Governance}
\label{app:construction}

\subsection{State Selection and Public Schema}

The builder begins from trajectories that expose an issue identifier, repository and base commit, typed tool actions, observations, opened files, search queries and results, test output when present, and a state update. Candidate boundaries are extracted before search, after search, after file inspection, and before an edit. A state enters Test500 only when the repository snapshot is reconstructable, the prefix contains genuine task progression, the information need belongs to the selected decision, observed source spans can be aligned, the content fingerprint is unique, near-duplicate prompts are removed, and the source instance belongs to the frozen Test500 inventory.

State fields are normalized without rewriting source text to simplify annotation. The public interface contains the issue, current information need, current observation, opened files, search queries, latest test output, current diff, hypothesis, subgoal, and visible trajectory prefix. Framework-specific fields can be retained in a namespaced metadata object. A stable state ID and content fingerprint allow another framework to emit an equivalent state card without reproducing the original private runtime.

\paragraph{Trajectory generation.}
Trajectories are produced by \texttt{adaptive\_llm\_trace\_v3\_agent}, an instrumented harness written for this benchmark so that every tool action, observed source span, and state update is recorded. For each issue the repository is materialized at its base commit and indexed into at most 1,600 provenance-preserving chunks served by path-aware BM25. The agent drafts a plan before it sees any repository content, issues two searches returning eight chunks each, reads up to three previously unopened files in windows of roughly 120 lines, and regenerates its retrieval intent, information need, hypothesis, and subgoal after every observation under the frozen \texttt{observation\_refresh\_v3} policy. One trajectory is run per issue and stops at the intended edit, which isolates the evidence need of a decision before any repair is committed; where a trajectory contains several searches or file reads, the last state matching each boundary type is kept. Test500 trajectories use DeepSeek-v4-Flash at temperature 0.1 with thinking disabled. Cal500 uses the same harness and refresh policy across a mix of planners: 399 states from a locally served Qwen3-Coder-30B, 99 from DeepSeek-v4-Flash, and two from a deterministic offline planner. Released state cards omit the planner identity, so no evaluated method can condition on it.

\paragraph{Source issues.}
The Test500 issue pool is frozen before any trajectory is generated: 362 candidate issues over 70 repositories, 320 of them from Multi-SWE-bench~\citep{zan2025multiswebench} and 42 from SWE-bench-Live~\citep{zhang2025swebenchlive}, each bound to a real repository and a fixed base commit, with a reserve held for replacements. Trajectory completeness, state quality, candidate-pool scorability, and certificate review reduce this to the released 500 states over 242 issue instances and 45 repositories. The pool is frozen without reading any method result, and it shares no repository with Cal500.

\paragraph{Cohort overlap.}
Action52 and Fresh23 follow the same harness, screening and certificate protocol as Test500 but are frozen as separate issue cohorts rather than drawn from the Test500 pool above, so the three evaluation cohorts are not repository-disjoint from one another. Candidate materialization and observed-evidence handling are cohort-specific and are documented with each evaluation. Action52 and Fresh23 share eight repositories, Action52 and Test500 five, and Fresh23 and Test500 six, with \texttt{lektor/lektor} the only repository common to all three. At the instance level the overlaps are eight, three and three, and \texttt{lektor\_\_lektor-1224} is the only instance shared by all three. The cohorts score different objects against different private certificates: a returned set against a sufficiency certificate, a repair-localization decision, and an executed test outcome.

\subsection{Calibration and Test Composition}

\begin{table}[ht]
\centering
\caption{Final SERBench composition.}
\label{tab:appendix-composition}
\small
\begin{tabular}{lrrr}
\toprule
Collection & Repositories & Issues & States \\
\midrule
Cal500 & 174 & 241 & 500 \\
Test500 & 45 & 242 & 500 \\
\bottomrule
\end{tabular}
\qquad
\begin{tabular}{lrr}
\toprule
State type & States & Graded \\
\midrule
Before search & 200 & 200 \\
After search & 124 & 124 \\
After file inspection & 104 & 104 \\
Before edit & 72 & 72 \\
\midrule
Total & 500 & 500 \\
\bottomrule
\end{tabular}
\end{table}

\Cref{tab:appendix-composition} gives the final composition. SERBench has two fixed partitions. Cal500 publishes 500 states and their certificates for development and calibration. Test500 publishes 500 test queries, while its certificates remain in the organizer evaluator and alone determine leaderboard results. The partitions are disjoint by state, issue instance, and repository. Reserve and excluded review records remain outside both released partitions.

Test500 certificates contain 1, 2, 3, 4, 5, and 6 required groups for 216, 135, 121, 22, 5, and 1 states, respectively, for a mean of 1.936 requirements per state. A majority of states, 284 of 500, require at least two independently grounded facts. The single-group stratum retains decisions whose unresolved need is atomic.

\subsection{Evidence Segmentation and Observed Alignment}

Repository snapshots are segmented into bounded evidence units. A unit stores a stable evidence and corpus ID, repository-relative path, line or structural span, symbol and evidence-type metadata when available, intact L2 text, compact L1 content, an L0 navigation card, token length, hashes, and reconstruction provenance. Units preserve one behavior or constraint while remaining independently retrievable. Overlapping chunks resolve to one canonical unit before scoring. When several source units independently support the same requirement, the certificate records them as alternatives instead of rewarding duplicates.

Observed intervals come from file reads and search results in the trajectory prefix. Exact path and overlapping line ranges are matched first. Normalized-content alignment handles tool output without precise spans. The aligned evidence IDs record $\observed$. Test500 evaluates the frozen candidate pools as released, including observed units where present, and prefix scores retain their original positions. Appendix~\ref{app:observed-overlap} reports performance by exact candidate/observation ID overlap. In a separate 21-state pilot that ranked a full repository index before any filtering, 53.3\% of a state-conditioned query's top ten repeated visible evidence, compared with 18.1\% for issue retrieval.

\subsection{Candidate Discovery and Task Boundary}

SERBench separates candidate discovery from evidence-set recovery. The public pool for each state is formed from a frozen union of gold-blind full-repository retrieval routes and source units discovered through the future trajectory, gold patch, and test patch. Units are canonicalized and construction-source tags are stripped before release. All reported methods use the same frozen candidate pool for each state; the pools retain recorded observed units in 186 states, enabling a direct robustness diagnostic. The organizer certificate is then grounded in this released pool, but certificate fields and source-origin labels are unavailable to retrieval methods. Fixing the candidates has two purposes. First, every pool contains a complete set, so a failure can be attributed to selection. Second, experts can check that a certificate lists every acceptable unit in a pool of 20 to 120 units, which is not feasible over a whole repository.

The primary task is candidate-pool evidence-set recovery: once candidate discovery supplies a recall-complete search region, the method must identify a set covering the decision's residual requirements. In the separate gold-blind full-repository diagnostic, scored against its own certificate snapshot, BM25 top-1000 covers a complete certificate for 66.6\% of states and 75.96\% of required groups. Qwen3 reranking raises complete-certificate coverage at depth 100 from 36.8\% to 61.8\%, while sharing the same 66.6\% top-1000 coverage ceiling. Within the released pools, an oracle reaches 99.80\% Complete-MSS@5. The remaining state requires six distinct units and has no admissible branch of at most five units. Uniform selection of five distinct units from each released candidate pool has an exact expected Complete-MSS@5 of 5.17\%. The released-pool values use the current Test500 certificates, group thresholds, and complete alternative branches.

\section{Grouped-MSS Annotation and Governance}
\label{app:annotation}

\subsection{Annotation Object}

Each private certificate conforms to \texttt{serbench-mss-v1}. Required fields identify the state, retrieval-needed status, evidence groups, release scope, and confidence. Every evidence group stores a semantic role, acceptable evidence IDs, a minimum required count, a necessity weight, a rationale, and source quotations. Optional fields record observed evidence, source-first repair history, and enumerated complete alternatives.

The schema represents two different forms of plurality. Interchangeable source units for one fact are recorded inside the same group, with the group's declared threshold $q_j$ determining how many acceptable IDs are needed. When $q_j=1$, the group is an OR over those IDs; larger thresholds require the stated count. Distinct complete solutions are stored in \texttt{alternative\_minimal\_sets}. Each entry is an AND branch, and satisfying any complete branch is sufficient. The scorer evaluates this nested structure directly rather than flattening the union of branch members, which would accept a mixture drawn from different solutions. In Test500 every enumerated branch is derived from the grouped requirements and satisfies them. The released scorer also accepts a recovered branch directly, so certificates whose branches are not derived from the groups remain scorable without a schema change.

Observed-evidence annotation follows the provenance of a claim rather than lexical overlap with the state card. File contents and search results captured by the trajectory, together with explicit \texttt{observed\_evidence\_ids}, count as acquired evidence. Agent-authored plans and hypotheses remain part of the query but require source confirmation. Reviewers first establish this epistemic boundary, then identify the unresolved requirements, ground them in source units, and test whether every retained requirement contributes nonredundant value to a concrete next search, edit, or verification action.

\subsection{Independent Review and Source-First Resolution}

Gold production and validation use two sequential safeguards. First, a cross-family model calibration audit has Grok~4.6 and Gemini~3.6~Flash independently inspect all 500 certificates without method predictions or human opinions; Claude Sonnet~5 provides source-constrained arbitration for 455 selected rows. All three reviewers are accessed through OpenRouter. These model outputs are review signals rather than automatic label changes, and \Cref{tab:model-audit-v01} records what each reviewer recommended. Second, a blinded human closure has two expert annotators independently review all 500 states, followed by a senior expert who arbitrates substantive anonymized A/B disagreements without receiving model decisions. Source-first human decisions are then applied to the release: every adjudicated certificate repair is incorporated and every inserted replacement is human-reviewed. This closure produces the final 500-state Test500 release and its aggregate governance ledger. Certificate changes require support from the visible state, repository source, quotations, and public candidate pool.

\begin{table}[ht]
\centering
\caption{Cross-family model audit recommendations before source-first adjudication.}
\label{tab:model-audit-v01}
\small
\setlength{\tabcolsep}{4pt}
\begin{tabular*}{\textwidth}{@{\extracolsep{\fill}}lrrrrrr@{}}
\toprule
Reviewer & Rows & Accept & Revise & Rebuild & Unscorable & Uncertain \\
\midrule
Grok 4.6 & 500 & 23 & 167 & 285 & 25 & -- \\
Gemini 3.6 Flash & 500 & 117 & 363 & 5 & 15 & -- \\
Claude Sonnet 5 & 455 & 93 & 54 & 1 & 53 & 254 \\
\bottomrule
\end{tabular*}
\end{table}

The different marginal distributions reveal distinct review thresholds, especially for whether a planning-state need is sufficiently grounded and whether a certificate is minimal. Source inspection separates genuine certificate plurality from four other causes of disagreement: the observed/unobserved boundary, quotation or provenance fidelity, joint sufficiency and minimality, and the action boundary represented by the state. Alternative branches resolve genuine plurality. Grounding and state defects instead require repair, reconstruction, or replacement.

Both independent human reviewers completed every assigned state. The four retrospective sensitivity
classes are S0 (no final change), S1 (a certificate edit with no Complete-MSS change for any locked
method at $k\in\{5,8,16\}$), S2 (a change to at least one such outcome), and S3 (replacement of a
state whose decision boundary is not independently scorable). A 200-state retrospective audit applies these classes. It is deliberately stratified rather than
drawn at random, so its class counts describe that audited sample and are not release-wide
rates, and it gives 55/26/82/37 cases for
S0/S1/S2/S3; a single flip at any locked cutoff assigns S2. Method outcomes were unavailable
during review.

\subsection{Final Certificate Accounting}

\Cref{tab:final-gold-v02} summarizes the released certificates. Across the complete human-audit closure, the ledger records 232 certificate-repair actions and 93 replacement decisions or mappings. These are governance events rather than mutually exclusive state counts: intermediate replacement queues and raw review flags are not final counts, while the final release contains 500 unique scorable states.

\begin{table}[ht]
\centering
\caption{Final source-reviewed Test500 certificates. Update counts are governance events and do not partition the 500-state denominator.}
\label{tab:final-gold-v02}
\small
\begin{tabular}{lr}
\toprule
Property & Count \\
\midrule
Scorable states & 500 \\
Independent model reviews & 1{,}000 \\
Third-model arbitration rows & 455 \\
Final human-audited states & 500 \\
Certificate-repair actions (cumulative) & 232 \\
Replacement decisions/mappings (cumulative) & 93 \\
States with enumerated complete branches & 454 \\
States with multiple complete branches & 124 \\
Enumerated complete branches & 955 \\
\bottomrule
\end{tabular}
\end{table}

\subsection{Independent Validity Audit of the Frozen Release}
\label{app:iaa}

This audit sampled 80 states uniformly at random from the 500-state release and assigned them to two
experts outside certificate construction and arbitration. They reviewed independently, without access
to each other's verdicts, and completed all 155 certificate groups in the sample. Each state received
an accept, revise, or rebuild verdict; each group received eight field-level judgments covering
requirement validity, source grounding, joint sufficiency, necessity, the minimum-required threshold,
acceptable-evidence completeness, observed-evidence contamination, and alternative-branch structure.

\begin{table}[ht]
\centering
\caption{Independent validity audit of 80 states drawn uniformly at random from the Test500 release,
by two experts outside the construction and arbitration chain.}
\label{tab:iaa}
\small
\begin{tabular}{lr}
\toprule
Property & Value \\
\midrule
States audited & 80 of 500 \\
Certificate groups audited & 155 \\
Expert A verdicts (accept / revise / rebuild) & 78 / 1 / 1 \\
Expert B verdicts (accept / revise / rebuild) & 79 / 1 / 0 \\
States accepted by both & 77 (96.25\%) \\
States flagged by at least one & 3 (3.75\%) \\
States flagged by both & 0 (0.00\%) \\
Group-level agreement on all eight fields & 151 of 155 (97.42\%) \\
Per-field agreement range & 97.42--100\% \\
Raw agreement (state / group) & 96.25\% / 97.42\% \\
$2p_o-1$ agreement (state / group) & 0.925 / 0.948 \\
\bottomrule
\end{tabular}
\end{table}

\Cref{tab:iaa} reports the audit in full. The auditors jointly accepted 77 of 80 states and gave identical judgments on all eight fields for 151 of 155 groups. No state was flagged by both, and both recorded confidence four or five on every state.
Cohen's $\kappa$ is uninformative at these margins: with the two auditors accepting 97.5\% and
98.75\% of states, expected agreement nearly equals observed agreement and $\kappa$ falls toward
zero despite agreement above 96\%. We therefore report raw agreement alongside the
prevalence-adjusted, bias-adjusted statistic $2p_o-1$.

All 155 groups received identical judgments on requirement validity, the minimum-required threshold,
observed-evidence contamination, and alternative-branch structure. The remaining disagreements were
on source grounding, joint sufficiency, and acceptable-evidence completeness, with every group-level disagreement belonging to one state. This judgment concerns the certificate's acceptable evidence, not whether the public candidate pool contains any observed units.

These are reported as agreement statistics for the audit, not as a further revision pass: the
released labels were unchanged.

After the freeze, deterministic checks validate 500 state/label joins, evidence IDs, recorded observed-evidence IDs, source quotations, branch membership, and public-pool provenance. A separate offline audit confirms that across all 500 frozen certificates no recorded observed ID appears in an acceptable set or in a complete alternative branch, and checks candidate/observed intersections without changing the frozen inputs or predictions. Initial reviews were blinded to method outcomes; targeted checks and replacement provenance remain in the audit record. Predictions are rescored against the release, with new retrieval outputs required only when public inputs change. All adjudicated repairs were applied, inserted replacements were human-reviewed, and the release passed structural, provenance, and scorer checks. The archive retains blinded reviews, source-grounded decisions, replacements, per-state provenance, and the retrospective severity analysis.

\section{Metrics and Baseline Implementations}
\label{app:metrics}

For state $t$, let $(e_1,\ldots,e_k)$ denote the ranked prediction. Group $g_j$ is covered when at least its declared threshold $q_j=\texttt{minimum\_required}$ acceptable IDs appear. Thus $q_j=1$ is an OR over acceptable evidence IDs; a group with $q_j>1$ requires the stated count. Complete-MSS accepts the conjunction of all required groups or one fully contained enumerated alternative branch. Each alternative branch is itself a complete AND solution, and the scorer never flattens distinct branches into a union. Group recall averages group coverage within each state before macro-averaging across states. Necessity-weighted recall replaces the uniform group average with normalized annotation weights. Grouped nDCG adds gain only when a newly covered group first appears.

The public scorer accepts JSONL rows containing a state ID, method ID, and ordered evidence IDs. The scorer joins each frozen ranking to its state certificate inside the organizer evaluator and evaluates the original first $k$ positions. An observed ID is not removed or replaced by a later item. The release audit verifies all 500 state/certificate joins, nonempty certificates, valid public evidence IDs, and separation between predictions and private labels.

\subsection{Common Evaluation Protocol}
\label{app:evaluation}

Every method is evaluated on the same public state card and frozen candidate pool; Test500 certificates remain in the organizer evaluator until predictions are complete. The primary \msscomplement configuration is selected on Cal500 and applied to Test500, the gold-blind full-repository diagnostic, AMA-208, Prospective44, and Action52 without retuning. Ablations and controller-family comparisons vary only their declared factors.

SERBench intervals use 20,000 paired repository-level bootstrap resamples clustered by repository. AMA intervals use 20,000 episode-level resamples with seed 20260901, and Action52 intervals use 10,000 task-level resamples with seed 20260904. The observed-overlap diagnostic uses seed 20260915 and 20,000 paired repository-cluster resamples within each stratum, weighting states equally; the reported percentile intervals are descriptive. Semantic calls use temperature zero and structured outputs. DeepSeek calls use DeepSeek's official API; Claude, Gemini, and Grok calls used for executor comparison or adjudication use OpenRouter; open-weight models, including retrievers and the Qwen3.6-35B-A3B judge, run locally.

\subsection{Complete Test500 Baseline Matrix}
\label{app:baselines}

All retrievers rank the complete released candidate pool. General retrievers choose between issue-only and structured state-plus-need queries on a fixed 75-state subset of Cal500. Agent-aware retrievers choose between their supported reasoning or information-need representations on the same subset. These choices are fixed before Test500 evaluation.

BM25~\citep{robertson2009bm25} uses the standard probabilistic score over the frozen query and source units. BGE-large~\citep{xiao2023bge} ranks by dense cosine similarity. SPLADE++~\citep{formal2022spladepp} uses learned sparse vectors, and ColBERTv2~\citep{santhanam2022colbertv2} uses late interaction. ReasonIR-8B, SweRankEmbed-Large, and Qwen3-Embedding-8B use their released similarity functions~\citep{shao2025reasonir,reddy2026swerank,zhang2025qwen3embedding}. The Qwen3 reranking row retrieves 100 units with Qwen3-Embedding-8B and reranks them with Qwen3-Reranker-8B. AgentIR-4B~\citep{chen2026agentir} receives the visible reasoning state and current need. Agent-ModernColBERT~\citep{chaffin2026agentmoderncolbert,chen2026agentir} uses its agent-aware late-interaction representation. The reported Test500 rankings are evaluated without an observed-evidence filter.

\paragraph{Released checkpoints.}
\Cref{tab:checkpoints} gives the exact model repository behind each evaluated baseline.

\begin{table}[ht]
\centering
\caption{Evaluated retrieval checkpoints. Each is pinned to a fixed revision in the released
configuration contracts.}
\label{tab:checkpoints}
\small
\begin{tabular}{ll}
\toprule
Baseline & Checkpoint \\
\midrule
BGE-large & \texttt{BAAI/bge-large-en-v1.5} \\
SPLADE++ & \texttt{prithivida/Splade\_PP\_en\_v1}, \texttt{Qdrant} ONNX conversion \\
ColBERTv2 & \texttt{colbert-ir/colbertv2.0} \\
ReasonIR-8B & \texttt{reasonir/ReasonIR-8B} \\
SweRankEmbed-Large & \texttt{Salesforce/SweRankEmbed-Large} \\
Qwen3-Embedding-8B & \texttt{Qwen/Qwen3-Embedding-8B} \\
Qwen3-Reranker-8B & \texttt{Qwen/Qwen3-Reranker-8B} \\
AgentIR-4B & \texttt{Tevatron/AgentIR-4B} \\
Agent-ModernColBERT & \texttt{lightonai/Agent-ModernColBERT} \\
\bottomrule
\end{tabular}
\end{table}

The MMR row starts from the Qwen3 reranker and freezes its relevance--diversity coefficient on calibration data. The selected coefficient is 1.0, recovering the original ranking. The independent-similarity control of \Cref{tab:attribution-budget} begins from the same nine-view fused ranking as \msscomplement. Its per-call geometry is as follows. The Pro call independently selects up to eight units from the first 80. A Flash call independently selects up to 48 from the first 120 without seeing or excluding that first selection. A final Flash call reranks their union against the frozen fused order and returns exactly seven units.

Displayed differences are calculated from unrounded values and then rounded, so subtracting rounded table entries can differ in the last decimal place.

\paragraph{State-set reranker (SSR).}
SSR is our one-pass semantic set reranker. It reads the visible state and a retrieved candidate union, then selects five evidence IDs with structured-output validation and deduplication. On Prospective44, the frozen report uses DeepSeek-v4-flash, one call, the leading ten candidates from each of two BGE-refined lexical routes, 1,600-character evidence excerpts, and a 2,400-token response limit. These limits describe controller input excerpts and response generation, not a universal returned-source-token budget. The AMA adapter instead uses trajectory-memory candidates, a 32-card semantic shortlist drawn from a 160-candidate route pool, one DeepSeek-v4-flash call, and a 4,096-token output-context ceiling. Cohort manifests retain these adapter settings and the corresponding predictions.

\subsection{Complete Budget Curves}

\begin{table*}[htbp]
\centering
\small
\caption{Complete-MSS across prefix caps on frozen Test500 predictions. A row remains constant after its available output ends.}
\label{tab:complete-attribution-curves}
\setlength{\tabcolsep}{6pt}
\begin{tabular*}{\textwidth}{@{\extracolsep{\fill}}lrrrrrrr@{}}
\toprule
Method & $k=1$ & $k=3$ & $k=5$ & $k=8$ & $k=10$ & $k=12$ & $k=16$ \\
\midrule
Qwen3 + Reranker & 27.8 & 48.4 & 61.4 & 72.4 & 77.0 & 80.0 & 85.8 \\
Independent similarity (fixed 7) & 28.4 & 53.4 & 66.6 & 74.4 & 74.4 & 74.4 & 74.4 \\
Single-stage MSS proposal & 24.8 & 53.4 & 70.0 & 79.4 & 79.4 & 79.4 & 79.4 \\
\msscomplement & 27.0 & 58.8 & \textbf{73.0} & \textbf{80.6} & 80.6 & 80.6 & 80.6 \\
Fixed top-8 finalization & 27.0 & 58.8 & 73.0 & 81.4 & 81.4 & 81.4 & 81.4 \\
\bottomrule
\end{tabular*}
\end{table*}

\Cref{tab:complete-attribution-curves,tab:fair-source-budget} report the full prefix curves and matched budgets. At sixteen items, Qwen reaches 85.80\% Complete-MSS with 8,643 mean source tokens. The matched item and token budgets isolate recovery within a common delivery constraint.

\begin{table}[htbp]
\centering
\small
\caption{Fair source-budget comparisons against Qwen3 embedding plus reranking. Values are Complete-MSS percentages; context is mean source tokens actually admitted, which falls below the returned mean in token-capped rows because individual states can exceed the cap.}
\label{tab:fair-source-budget}
\setlength{\tabcolsep}{5pt}
\begin{tabular*}{\textwidth}{@{\extracolsep{\fill}}lrrrr@{}}
\toprule
Condition & Ours & Qwen & $\Delta$ & Ours/Qwen context \\
\midrule
5 items & 73.00 & 61.40 & +11.60 & 2,543 / 2,721 \\
8 items & 80.60 & 72.40 & +8.20 & 3,663 / 4,323 \\
4,096 tokens & 80.00 & 71.20 & +8.80 & 3,307 / 4,056 \\
6,144 tokens & 80.60 & 80.00 & +0.60 & 3,655 / 6,100 \\
Per-state matched & 80.60 & 69.20 & +11.40 & 3,663 / 3,628 \\
\bottomrule
\end{tabular*}
\end{table}

\subsection{Performance by Observed-Candidate Overlap}
\label{app:observed-overlap}

Released candidate pools retain source units the agent had already observed in some states
and not in others. This appendix partitions Test500 by that property and reports the main
retrieval comparison within each stratum. The diagnostic was conducted after the primary
analysis and aggregates the same frozen predictions without reranking, rescoring, or
retuning.

A state belongs to the overlap stratum when its released candidate pool contains at least
one evidence ID recorded in its observed set. The partition is read from the released
per-candidate observation flag and independently reproduced by an exact evidence-ID join;
the two agree on all 500 states and on all 953 flagged candidate entries. The strata are
mutually exclusive and exhaust the 500 states: 314 without recorded overlap and 186 with.
They cover 45 and 31 repositories, which do not sum to the cohort total because a
repository can contribute states to both. Overlap here means recorded exact evidence IDs,
not semantic or line-range disjointness.

\begin{table}[htbp]
\centering
\small
\caption{State composition of the two strata by decision boundary.}
\label{tab:observed-overlap-composition}
\setlength{\tabcolsep}{5pt}
\begin{tabular*}{\textwidth}{@{\extracolsep{\fill}}lrrrrr@{}}
\toprule
Stratum & Before search & After search & After file inspection & Before edit & Total \\
\midrule
No recorded overlap & 200 & 6 & 63 & 45 & 314 \\
Recorded overlap & 0 & 118 & 41 & 27 & 186 \\
\bottomrule
\end{tabular*}
\end{table}

The strata differ systematically in trajectory stage: every state with recorded overlap
follows a search, while most states without it precede one.

\begin{table}[htbp]
\centering
\small
\caption{Test500 main retrieval methods within each overlap stratum. Columns match
\Cref{tab:main-retrieval}; best value per column in bold.}
\label{tab:observed-overlap-recovery}
\setlength{\tabcolsep}{2.2pt}
\begin{tabular*}{\textwidth}{@{\extracolsep{\fill}}lrrrrr@{}}
\toprule
Method & Complete@5 & Complete@8 & Group@5 & Group@8 & Necessity@5 \\
\midrule
\multicolumn{6}{l}{\emph{No recorded overlap, $n=314$}} \\
BM25 & 10.51 & 15.92 & 15.50 & 22.49 & 15.61 \\
BGE-large & 35.35 & 44.90 & 46.99 & 57.31 & 47.13 \\
SPLADE++ & 29.62 & 40.76 & 40.38 & 51.34 & 40.86 \\
ColBERTv2 & 18.15 & 26.11 & 25.16 & 34.34 & 25.32 \\
ReasonIR-8B & 34.08 & 43.95 & 44.08 & 55.94 & 44.21 \\
SweRankEmbed-Large & 55.10 & 64.97 & 68.82 & 77.37 & 69.33 \\
Qwen3-Embedding-8B & 54.46 & 66.24 & 67.46 & 77.69 & 67.52 \\
Qwen3-Embed + Reranker-8B & 65.92 & 77.39 & 79.09 & 86.94 & 79.83 \\
Qwen3-Embed + Reranker-8B + MMR & 65.92 & 77.39 & 79.09 & 86.94 & 79.83 \\
AgentIR-4B & 31.21 & 41.40 & 41.07 & 51.67 & 41.18 \\
Agent-ModernColBERT & 33.76 & 43.31 & 45.12 & 55.47 & 45.68 \\
\msscomplement & \best{74.20} & \best{81.85} & \best{83.99} & \best{89.74} & \best{84.48} \\
\addlinespace[2pt]
\multicolumn{6}{l}{\emph{Recorded overlap, $n=186$}} \\
BM25 & 6.45 & 8.06 & 9.09 & 11.56 & 9.02 \\
BGE-large & 23.66 & 33.87 & 30.33 & 42.56 & 30.21 \\
SPLADE++ & 13.44 & 20.97 & 17.61 & 27.02 & 17.53 \\
ColBERTv2 & 14.52 & 20.43 & 19.13 & 26.61 & 19.27 \\
ReasonIR-8B & 17.74 & 26.88 & 23.97 & 34.50 & 24.07 \\
SweRankEmbed-Large & 47.85 & 58.06 & 56.05 & 65.99 & 56.34 \\
Qwen3-Embedding-8B & 49.46 & 60.75 & 57.93 & 69.09 & 58.02 \\
Qwen3-Embed + Reranker-8B & 53.76 & 63.98 & 61.25 & 72.09 & 61.34 \\
Qwen3-Embed + Reranker-8B + MMR & 53.76 & 63.98 & 61.25 & 72.09 & 61.34 \\
AgentIR-4B & 51.08 & 57.53 & 58.38 & 66.04 & 58.73 \\
Agent-ModernColBERT & 34.41 & 42.47 & 42.11 & 51.70 & 42.37 \\
\msscomplement & \best{70.97} & \best{78.49} & \best{78.72} & \best{84.77} & \best{79.06} \\
\bottomrule
\end{tabular*}
\end{table}

\msscomplement ranks first on all five metrics in both strata, and Qwen3 embedding with
reranking is the strongest baseline in both.

\begin{table}[htbp]
\centering
\small
\caption{Paired difference between \msscomplement and Qwen3 embedding with reranking
within each stratum.}
\label{tab:observed-overlap-paired}
\setlength{\tabcolsep}{6pt}
\begin{tabular*}{\textwidth}{@{\extracolsep{\fill}}lrr@{}}
\toprule
Stratum & Complete@5 difference (pp) & Complete@8 difference (pp) \\
\midrule
No recorded overlap, $n=314$ & $+8.28$ $[+3.97,+13.31]$ & $+4.46$ $[+1.84,+9.14]$ \\
Recorded overlap, $n=186$ & $+17.20$ $[+8.02,+27.75]$ & $+14.52$ $[+8.26,+23.76]$ \\
\bottomrule
\end{tabular*}
\end{table}

Relative to Qwen3 embedding with reranking, the Complete-MSS gains at five and eight items
have paired 95\% confidence intervals entirely above zero in both strata. The main
retrieval advantage is therefore present in states both with and without recorded
observed-ID overlap in their candidate pools.

Intervals use 20,000 paired resamples clustered by repository within each stratum, with
seed 20260915; states are weighted equally and both methods use the same repository
multiplicities in each resample. Because the strata differ systematically in trajectory
stage, the difference between their effect sizes is descriptive rather than an estimate of
the causal effect of observed candidates. Weighting the two strata by state count
reproduces the full-cohort means in \Cref{tab:main-retrieval} exactly.

\subsection{Results by Required Evidence-Group Count}
\label{app:group-strata}

\Cref{tab:group-strata} separates the frozen Test500 states into mutually exclusive strata with one, two, and at least three required evidence groups. The strata count decision-specific requirement groups rather than distinct evidence items, since one source unit may satisfy several groups. Combining the two multi-group strata gives 284 states: the Complete-MSS gain over Qwen3 embedding plus reranking is 14.79 points at five items, $[+9.42,+22.61]$, and 10.56 points at eight, $[+6.76,+18.12]$. Both repository-clustered intervals exclude zero.

%% FROZEN TABLE DATA (v1.1; restored in v1.5).
%% Explicitly authorized relocation in v1.9 for the nine-page main-text target.
%% Preserve every row, column, value, and statistical meaning. The main text
%% retains the @5 stratified evidence; the complete @5/@8 table belongs here.
\begin{table}[htbp]
\centering
\caption{Test500 by the exact number of required evidence groups in the frozen certificate.
\msscomplement against Qwen3-Embedding-8B plus Qwen3-Reranker-8B, with 20,000 paired
repository-clustered bootstrap draws per row. Bold marks a difference whose 95\% interval excludes
zero. On single-group states with a unit threshold, Complete-MSS reduces to hit@$k$.}
\label{tab:group-strata}
\small
\setlength{\tabcolsep}{4.5pt}
\begin{tabular*}{\textwidth}{@{\extracolsep{\fill}}lrrrrrrr@{}}
\toprule
Required & & \multicolumn{3}{c}{Complete-MSS@5} & \multicolumn{3}{c}{Complete-MSS@8} \\
\cmidrule(lr){3-5}\cmidrule(lr){6-8}
groups & States & Qwen & \msscomplement & $\Delta$ & Qwen & \msscomplement & $\Delta$ \\
\midrule
1 & 216 & 78.70 & 86.11 & +7.41 & 85.19 & 90.28 & +5.09 \\
2 & 135 & 61.48 & 77.04 & \best{+15.56} & 75.56 & 85.93 & \best{+10.37} \\
$\ge 3$ & 149 & 36.24 & 50.34 & \best{+14.09} & 51.01 & 61.74 & \best{+10.74} \\
\addlinespace[2pt]
All & 500 & 61.40 & 73.00 & \best{+11.60} & 72.40 & 80.60 & \best{+8.20} \\
\bottomrule
\end{tabular*}
\end{table}

At five items the gains are 7.41 points for single-group states, $[-0.55,+14.39]$, 15.56 for two-group states, $[+7.14,+31.94]$, and 14.09 for states requiring at least three groups, $[+6.25,+21.67]$, where completion rises from 36.24\% to 50.34\%. At eight items, the gains are 5.09 points for single-group states, $[-0.81,+11.69]$, 10.37 for two-group states, $[+4.49,+23.00]$, and 10.74 for states requiring at least three groups, $[+5.60,+16.84]$. The two multi-group strata retain positive intervals at this larger prefix.

\subsection{Gold-Blind Full-Repository Diagnostic}
\label{app:fullrepo}

%% PLACEMENT: this float is declared before D.5's opening paragraph on purpose.
%% A_appendix.tex puts a \FloatBarrier after A3, so a table still pending when the
%% barrier arrives forces the page out and leaves the rest of it empty. Measured, not
%% guessed: declared after the paragraph the appendix is 30 pages with 500pt unused on
%% p21; declared here it is 29 pages with no hole. No wording changes either way.
%% FROZEN TABLE SPECIFICATION (v1.1). The metric set is deliberate: do not drop
%% columns or rows. Moved to the appendix in v2.10 by author decision. Section 6.4
%% states both contrasts with their intervals and the two token counts, and this
%% subsection already carried the row-by-row pointer.
\begin{table}[ht]
\centering
\caption{Gold-blind full-repository diagnostic on the fixed discovery cohort. Coverable denotes states whose complete certificate occurs in the BM25 top 1,000. Values are percentages, and the last column is mean source tokens in the five-item prefix.}
\label{tab:fullrepo-main}
\small
\setlength{\tabcolsep}{3.5pt}
\begin{tabular*}{\textwidth}{@{\extracolsep{\fill}}lrrrrrrrrr@{}}
\toprule
& \multicolumn{4}{c}{All 500} & \multicolumn{4}{c}{Coverable 333} & \\
\cmidrule(lr){2-5}\cmidrule(lr){6-9}
& \multicolumn{2}{c}{Complete} & \multicolumn{2}{c}{Group recall} & \multicolumn{2}{c}{Complete} & \multicolumn{2}{c}{Group recall} & Source \\
\cmidrule(lr){2-3}\cmidrule(lr){4-5}\cmidrule(lr){6-7}\cmidrule(lr){8-9}
Method & @5 & @8 & @5 & @8 & @5 & @8 & @5 & @8 & tok.@5 \\
\midrule
Qwen3 + Reranker & 33.80 & 40.80 & 50.69 & 55.69 & 50.75 & 61.26 & 67.88 & 74.62 & 4,290 \\
\msscomplement & \best{38.80} & \best{43.20} & \best{54.24} & \best{57.87} & \best{58.26} & \best{64.86} & \best{73.31} & \best{77.77} & \best{4,014} \\
\bottomrule
\end{tabular*}
\end{table}

The diagnostic retrieves from each frozen repository using only the visible state and observed mask. BM25 supplies the top-1,000 discovery pool. Qwen3-Embedding-8B ranks within that pool, and Qwen3-Reranker-8B reranks its leading 100 candidates. Private certificates enter only after all rankings and controller outputs are frozen. A complete certificate is present for 333 states in the fixed discovery cohort. This cohort carries its own certificate snapshot and shares 434 state IDs with the current Test500, and \Cref{tab:fullrepo-main} scores all 500 states against that snapshot. The all-state score composes discovery with selection, and the coverable score conditions on successful discovery.

For a common standard with Test500, we rescore the frozen predictions for the 434 shared states, from 39 repositories, against the current Test500 certificates. Each predicted unit maps to a Test500 unit only when the state, case-sensitive source path, line span, and SHA256 hash all match exactly, and unmapped units keep their rank positions without certificate credit. The visible states of all 434 are identical in both cohorts, the snapshots differ in certificate structure for 126 of them, and rescoring under the original snapshot reproduces every original per-state score. Under the current certificates, the BM25 top 1,000 contains a complete certificate for 308 of the 434 states, against 292 under the original snapshot. Over all 434 states, \msscomplement completes 44.47\% at five units against 40.55\% for Qwen3 + Reranker, a gain of 3.92 points, $[+1.56,+8.03]$. At eight units the gain is $+1.15$ points, $[-2.16,+7.14]$. On the 308 coverable states, \msscomplement completes 62.66\% at five units against 57.14\%, $+5.52$ points, $[+2.44,+11.94]$, and the eight-unit gain is $+1.62$, $[-3.35,+10.57]$. On the released pools, the same 434 states give 71.20\% against 58.76\%. Under the same certificates as Test500, finding candidates remains the main bottleneck, and set selection keeps a significant five-unit gain.

\subsection{Public Calibration Example}

Consider a Cal500 state from \texttt{arrow-py/arrow} immediately before an edit. The agent has already inspected 53 source units and must decide how to replace \texttt{dateutil} timezone behavior while preserving every call site. The public certificate has two required groups: the parsing behavior in \texttt{arrow/parser.py} and the locations that construct or consume timezone objects. Qwen3 embed-plus-reranker fills its top five with call-site evidence and covers only the second group. \msscomplement includes the parser implementation together with representative call sites, completing both groups under the same five-item limit.

\FloatBarrier
\section{MSS-Complement Operational Details}
\label{app:method-details}

\subsection{Frozen Candidate and Controller Interface}

This interface instantiates the proposal, expansion, and finalization calls in \Cref{fig:mss-complement-pipeline}. Source cards are the controller's views of candidate units, while the final selected units determine the downstream working context.

On the supplied candidate pool, nine label-hidden views rank intact source units: BM25 over the question, the full state, prior actions, and observations, word and character TF--IDF, dense similarity, entity overlap, and recency. Reciprocal-rank fusion admits the leading 64 units from each view and caps the union at 384. The first semantic call sees 80 source cards capped at 760 characters. The second and third calls each see at most 120 cards, capped at 280 and 620 characters. The second-stage expansion region contains 48 units. These are controller-card character budgets; the final working context is separately constrained to 6,144 source tokens.

The first call uses \texttt{deepseek-v4-pro} and the two revision calls use \texttt{deepseek-v4-flash}. Both aliases are called through DeepSeek's official API at temperature zero, with structured JSON outputs, a 240-second timeout, and at most three attempts, and were accessed during August--September 2026. Because the provider exposed no immutable revision identifiers, the artifact records aliases, access dates, prompts, and response hashes. The interface requires one state-interpretation controller followed by two selected-set revision calls. On Test500 the three calls average 17.65K, 6.69K, and 17.15K tokens per state, counting the logical prompt and completion tokens of each call, including calls answered from a response cache. At DeepSeek's off-peak list prices on 10 November 2026, without cache discounts, these calls cost about 1.5 US cents per state. Off-peak rates per million tokens are \$0.66 for input and \$1.98 for output on the first call's model and \$0.15 and \$0.60 on the model of the two later calls, and rates in peak hours are double. The first call, on the larger model, accounts for about three quarters of this cost, and the two later calls together cost about 0.4 cents.

Stage one proposes eight evidence IDs that jointly support a precise answer to the current information need. Stage two receives that proposal and deeper candidates, then identifies evidence for answer facts not yet supported. Stage three checks the combined set, revises the set to reduce repeated support, and returns 4--8 ranked source units. Unknown and duplicate IDs are removed. Deterministic validation enforces membership in the public candidate set and a 6,144 source-token ceiling. A deterministic ranking fallback is available if structured output fails; no reported final run used it.

The prompts never include a grouped-MSS certificate, group count, gold evidence ID, answer, or reward. The second and third prompts do see the currently selected set. This selected-set feedback conditions later choices on support already retained, while the candidate pool, scorer, and private labels remain fixed outside the calls. The selected set here is the current proposal, not the trajectory's observed set $\observed$; the Test500 adapter applies no additional filter based on recorded observed IDs.

\subsection{Attribution and Ablation Accounting}

The three-call variants in \Cref{tab:attribution-budget} use the same candidate generator, controller allocation, stage caps, source ceiling, decoding settings, and direct provider endpoint. The independent-similarity row removes the MSS objective and selected-set feedback. Of 1,500 controller responses, 1,498 have the state-specific requested length and 1,488 require no normalization; 12 responses are deterministically normalized against the frozen candidate relevance order, and 19 shorter second-stage lists are valid because those states have fewer than 48 candidates. Every final trace satisfies the stage cardinalities $\langle 8,\min(48,N),7\rangle$ and returns seven valid IDs in all 500 states. Excluding the 12 affected states leaves a Complete-MSS advantage of 5.74 points at five items, with a repository-clustered 95\% interval of $[+2.40,+10.22]$.

The two no-feedback rows are separate three-stage runs with inter-stage visibility disabled. Their objective, controller allocation, route depth, candidate pool, stage caps, card budgets, fusion, and 6,144-token ceiling are unchanged; stages two and three read static ranked prefixes, and the three proposals are fused by reciprocal-rank fusion. One fixes the final count at seven; the other adapts it between four and eight, returning four, five, six, seven, and eight units for 17, 31, 70, 85, and 297 states, respectively. Both cover 500 states over 1,500 logical calls with a recorded configuration hash, and no row in the table has private-label access at inference.

The derived rows are reconstructed offline from the corresponding frozen main-method traces. Single-stage uses the stage-one \texttt{initial\_ids} in recorded order and inherits nothing from the final-stage ordering. Against this row, the two later calls add 3.00 points of Complete-MSS@5 and return 476 fewer source tokens per state, $[-589,-327]$. At the list prices above, the two later calls account for about a quarter of the per-state cost, because both run on the smaller model. Fixed top-8 uses the complete final ranking and preserves its first five units. With eight units each, it is 2.00 points above the single-stage row on the full returned set, $[+0.27,+3.85]$, and the 4--8 output of the full policy is 0.80 points below it, $[-1.57,0.00]$, with 11\% less context. The $\lambda=1.0$ MMR row uses the corresponding Qwen ranking. These derivations make no additional API calls. All 16 released methods have 500 unique states and valid candidate IDs, and the consistency audit verifies cached completion, group-recall, and necessity-recall scores against the same current certificate file at every reported cutoff.

The no-expansion row reuses each state's frozen proposal $S_0$ and merged ranking and reruns only the finalization call. Its candidates are $S_0$ followed by the ranked reserve, deduplicated, capped at 120, and tagged by origin. The call keeps the original prompt, 4--8 output range, 6,144-token ceiling, and deterministic repair rules. All 500 calls returned valid output, those rules repaired or filled 65 selections without further calls, and the released scorer reproduces the resulting per-state scores. The run called \texttt{deepseek-v4-flash} in November 2026, after the original traces, so it is not contemporaneous with the other rows. It returns four, five, six, seven, and eight units for 143, 64, 76, 70, and 147 states, against 50, 35, 34, 58, and 323 for the full policy, a mean of 6.03 units against 7.14. Complete support is present in $S_0$ for 397 states and in its union with the expansion candidates for 456, a diagnostic without the final item limit. In 27 of the 29 states that expansion recovers on the full returned set, $S_0$ was already complete. Among the 24 states that the full policy completes at five units and the proposal does not, $S_0$ already held a complete set of at most five units in 18. Its union with the expansion candidates first held one in 4 more, the ranked reserve supplied the last 2, and the final five units come only from $S_0$ in 15 of the 24. In the 9 states that the proposal completes at five units and the full policy does not, expansion units entered the final five in 5.

\subsection{Post-Freeze Parameter Neighborhood}

The formal neighborhood changes route depth or candidate-pool width while keeping the prompt, controller allocation, scorer, and Cal120 state set fixed. Seven earlier compatible runs provide a broader neighborhood. They share the same Cal120 states, three-call family, scorer, and label-hidden inference, but may change more than one geometric or prompt factor; the strict rows isolate individual factors. \Cref{tab:cal120-robustness} reports both cohorts.

\begin{table*}[htbp]
\centering
\small
\caption{Cal120 post-freeze robustness. Strict rows change one declared factor; broad rows reuse compatible historical configurations. Ret.\ is returned completion and Nec.@5 necessity-weighted recall at five items. Values are percentages except context and seconds.}
\label{tab:cal120-robustness}
\setlength{\tabcolsep}{4.5pt}
\begin{tabular}{llrrrrrrr}
\toprule
Setting & Config. & Comp.@5 & Comp.@8 & Ret. & Group@5 & Nec.@5 & Context & Sec. \\
\midrule
Frozen center & strict & 65.00 & 75.83 & 75.83 & 79.65 & 79.64 & 3,178 & 6.315 \\
Route depth 48 & strict & 62.50 & 73.33 & 73.33 & 76.39 & 76.58 & 3,217 & 6.470 \\
Candidate pool 320 & strict & 64.17 & 74.17 & 74.17 & 79.24 & 79.42 & 3,171 & 4.877 \\
Broad A & historical & 65.00 & 74.17 & 74.17 & 77.22 & 77.52 & 3,362 & 4.725 \\
Broad B & historical & 68.33 & 75.00 & 75.00 & 79.93 & 80.16 & 3,217 & 5.017 \\
Broad C & historical & 65.83 & 72.50 & 72.50 & 80.49 & 80.68 & 3,214 & 5.630 \\
Broad D & historical & 66.67 & 74.17 & 74.17 & 80.76 & 81.11 & 3,185 & 5.218 \\
Broad E & historical & 65.00 & 74.17 & 74.17 & 78.75 & 78.91 & 3,262 & 5.127 \\
Broad F & historical & 60.83 & 68.33 & 68.33 & 75.35 & 75.59 & 3,234 & 5.153 \\
Broad G & historical & 62.50 & 71.67 & 71.67 & 76.74 & 77.18 & 3,233 & 5.454 \\
\bottomrule
\end{tabular}
\end{table*}

The strict neighbors preserve the qualitative operating point: complete-set performance changes by at most 2.5 points at five and eight items, with similar context. The broader neighborhood spans 60.83--68.33\% at five items and 68.33--75.00\% at eight. Repository-clustered paired intervals and the exact configuration mapping are included in the artifact ledger.

\subsection{Identity and Leakage Audit}

The frozen Test500 prediction file contains 500 unique state IDs and 1,500 logical controller calls. Every row records \texttt{labels\_read\_at\_inference=false}, with controller and parsing fallback counts at zero. The current offline audit recomputes all reported retrieval cutoffs against the frozen labels and checks the source traces used by the derived controls. The contract records configuration, implementation, public-input, private-label, and prediction hashes. A historical network-disabled replay additionally reproduced its corresponding source-run prediction file byte for byte. That replay establishes the source implementation, and the offline audit above covers the final merged release.

\FloatBarrier
\section{Supporting External and Action Evidence}

The following evaluations examine the evidence-acquisition objective beyond the main candidate-pool test. Prospective44 checks the configured policy on new states. AMA-Bench measures answering with the selected working context and accounts for the computation used to construct it. The action evaluation measures next-step repair localization from the supplied evidence, and Fresh23 carries the comparison through to executed repository tests.

\subsection{Prospective Confirmation}
\label{app:retrieval}

Prospective44 uses the same state schema and grouped-MSS scoring definition, with its own frozen candidate construction. Its 44 states were collected after the freeze and were not used for method selection. At five items, \msscomplement completes 20/44 states (45.45\%) versus 18/44 (40.91\%) for SSR; its group and necessity-weighted recalls are both 46.21\%. The mean returned context is 1,155 source tokens, and all 132 controller calls complete without fallback or private-label access.

\subsection{AMA-Bench Controlled Accounting}
\label{app:ama-accounting}

The controlled AMA comparison is a 208-episode real-world evaluation with 12 questions per episode. \msscomplement runs the Cal500 configuration with no AMA-specific tuning, so all 208 episodes are evaluated out of sample for the method. Every row uses DeepSeek-v4-flash for answer generation and a locally deployed Qwen3.6-35B-A3B for correctness judging. The complete token accounting is in \Cref{tab:external-main}. The AMA paper reports 57.22\% for AMA-Agent under its original generation and judging stack \citep{zhao2026amabench}; the shared stack used here gives 53.97\%, so the two values use different generation and judging stacks.

For the merged \msscomplement run, the answer and online totals include cumulative billed usage from the retained source runs, so the online column is an acquisition cost rather than the cost of one answer. The complete paired result is 359 gains, 1,830 ties, and 307 losses relative to AMA-Agent. The accuracy difference is $+2.08$ points with an episode-clustered 95\% interval of $[-0.16,+4.37]$. The answer prompt is 76.2\% smaller. Recorded full-run retrieval latency is 4.75 seconds per question for \msscomplement and approximately 6.94 seconds for AMA-Agent, about 31.5\% lower in this evaluated deployment. These wall-clock values are deployment measurements; the efficiency statement concerns the answer-visible prompt, while the acquisition tokens that produce it remain in the online-total column.

\subsection{Controller-Family Confirmation}
\label{app:controllers}

The controller-family comparison holds the 44 states, candidate input,
three-stage prompt semantics, evidence limits, and grouped-MSS scorer fixed.
Only the controller model family changes. Table~\ref{tab:controller-family}
reports the complete accounting.

\begin{table}[htbp]
\centering
\small
\caption{Prospective44 controller-family confirmation. Source context and
controller tokens are means per state. Latency is provider-dependent.}
\label{tab:controller-family}
\setlength{\tabcolsep}{4pt}
\begin{tabular*}{\textwidth}{@{\extracolsep{\fill}}lrrrrr@{}}
\toprule
Controller & Complete@5 & Group@5 & Context & Ctrl. tok. & Fallback \\
\midrule
DeepSeek & 45.45 & 46.21 & 1,155 & 43,696 & 0 \\
Claude & \textbf{52.27} & \textbf{53.03} & 1,168 & 63,999 & 0 \\
\bottomrule
\end{tabular*}
\end{table}

Each run covers 44/44 states. Neither run reads organizer-private
certificates during inference. The
Claude run costs \$5.78 through OpenRouter. Recorded mean retrieval latency is
6.79 seconds for DeepSeek and 16.07 seconds for Claude, but the provider,
serving path, and billing endpoint differ, so this number is not used to rank
the controller families.

\FloatBarrier
\subsection{Action52: Retrieval to a Scorable Downstream Decision}
\label{app:action52}

\paragraph{Purpose and task construction.}
Action52 evaluates whether evidence-set recovery is useful for completing
the next agent task. It separates three objects: the evidence returned by a
retriever, the production-code targets proposed by an executor, and the
reference files used by an offline evaluator. A retrieval score is therefore
not reused as the downstream outcome. The executor must interpret the
retrieved material and produce its own repair-location decision.

The cohort contains 52 task instances from 52 repositories. Each has an intermediate state, a source-grounded certificate, and a reference
patch that passes executable preflight. For 32 cases, the target
files come from a constructed minimal residual reference repair. For the
other 20, they come from the retained files in the validated historical
reference patch. These references define the file set scored for the next repair.
They are retained by the organizer and are not shown as candidate answers.
The main analysis retains all 52 cases.

\paragraph{Reference scope.}
The reference set contains 156 file entries, and the reported endpoint is coverage of that reference set rather than of production code alone. Per-case reference sizes and path composition are preserved in the accompanying audit.

\paragraph{Evidence conditions and budgets.}
Both learned retrievers receive the materialized residual evidence pools for the independent Action52 states, with observed evidence excluded. They do not perform a new full-repository discovery run. Action52 uses its own certificates and declared returned-set budgets, so its completion percentages are not an additional Test500@5 result.
Qwen3-Embedding-8B uses the structured-state query without a
reranker, and the reranked condition applies Qwen3-Reranker-8B to that
embedding's leading 100 candidates. \msscomplement uses the final retrieval output. The three retrieval
conditions contain the same number of evidence units for every case,
between five and eight, with 395 units across the 52 cases in each condition.
The evidence renderer enforces a 6,144-source-token ceiling per case.
Observed evidence is excluded. Identical materialized bundles are used by
both downstream executors.

Oracle-MSS is constructed by the organizer from the certificate.
It selects an admissible source unit for each required group, deduplicates
shared units, and fills the remaining slots from the Qwen ranking.
One case needs one more unit than the adaptive method returned to preserve
complete coverage. The oracle therefore contains 396 units in total and
still obeys the eight-unit ceiling. The executor sees the source material,
not group labels, a certificate, or a reference patch.

The Oracle-minus-one-required-group condition removes the group with greatest
necessity weight, breaking ties by group identifier. All registered alternatives for
that group are excluded, and the vacated slots are filled from the
ranking with other source units. It preserves the Oracle's per-case item
count and token ceiling and is supplied unchanged to both executor families.
Complete-MSS is zero by construction while mean group recall stays at 48.27\%, which is what makes this a paired missing-required-group control with substantial evidence still in place.

For a certificate containing $g$ required groups, let $h_j$ indicate whether
the supplied evidence satisfies group $j$. In Action52 each group requires
one admissible source unit. Group recall is $g^{-1}\sum_{j=1}^{g}h_j$, while
Complete-MSS is $\prod_{j=1}^{g}h_j$. A three-group case with two groups
recovered therefore scores $2/3$ on group recall and zero on Complete-MSS.
We average each score over tasks rather than pooling groups across tasks.
These are certificate-based scores. Because overlapping source windows can carry the same fact under different identifiers, the removal check inspects the materialized source content and the shared state alongside the removed identifiers.
For four confirmed duplicate-window cases, the removal also excludes
overlapping target implementation spans and matching implementation
signatures before filling the vacated slots. The original target group and
all other groups are preserved, and both executors use the corrected views.
Their reference-file scores are unchanged after this correction. Public
state clues, including an issue's existing repair suggestion, remain shared
across conditions. The accompanying source audit retains the remaining partial-overlap flags, so the scope of each removal is recoverable case by case.

\paragraph{Executor contract.}
The executor receives the repository name, public state, evidence text and
identifiers, and a target-count cap. It is asked to choose production-code
targets to inspect for the next repair step and returns JSON containing
relative paths, optional symbols, a short intent, invariants, and evidence
citations. It may propose a new production file. The prompt instructs the
executor not to choose tests, fixtures, snapshots, CI files, or evaluator
files. Paths are generated freely, with no answer menu, source-window lookup, repository search, patch generation, or test feedback in this run.

For a reference repair with $r$ retained files, the common target cap is
$\min(6,\max(3,r))$. This organizer-derived scalar is supplied equally to all conditions, while reference paths and patch contents remain hidden. Path
normalization and path filtering are deterministic. Extra paths consume the cap without incurring a separate precision penalty. An empty
valid selection receives zero recall and zero success.

DeepSeek-v4-flash uses the direct API and Claude Sonnet 5 uses OpenRouter.
Both use temperature zero, a 900-token response limit, and the same decision
prompt and JSON contract. Up to three format attempts request a corrected
or shorter JSON object without adding task evidence. Every condition has
52 final case records, with no provider substitution. The DeepSeek
missing-group condition contains one terminal empty selection, retained as
a zero-score case. The experiment changes the downstream executor family while holding the retrieval outputs fixed, which keeps it distinct from the retrieval-controller experiment in Appendix~\ref{app:controllers}.

\paragraph{Target-level adjudication and equal-budget scoring.}
The executor may return fewer targets than its cap, so the number of emitted
targets is not forced to be equal after inference. Every emitted target is
reviewed by a human against the source, state, and approved repair-localization
scope. A \emph{precise} target identifies a correct and sufficiently local
function, branch, statement, or insertion relation. A \emph{coarse} target is
related but does not isolate an actionable location; a \emph{wrong} target is
unsupported or irrelevant. Six proposals were explicitly adjudicated as not
precise without forcing a finer coarse-versus-wrong label.

Traditional target precision divides precise targets by the number actually emitted, and its denominator ranges from 161 to 169 across conditions. \Cref{tab:action52-target-detail} retains it for reference. The main table
instead reports \emph{fixed-budget target precision}: the number of precise targets
divided by the sum of the predeclared per-case caps. Because all conditions
use the same 52 cases, this denominator is 171 for every row. An unused slot
receives zero credit. This preserves the executor's decision to return fewer
targets without deleting, padding, or relabeling any generated proposal.

\paragraph{Deterministic endpoints.}
For case $i$, let $R_i$ be the organizer's reference-file set and
$P_i$ the executor's valid predicted paths. Reference-file recall is
$|P_i\cap R_i|/|R_i|$, averaged equally across all 52 cases. Localization
success is the indicator that $R_i\subseteq P_i$. Complete-MSS and group
recall are computed separately from the supplied evidence using the
certificate. Optional symbols and target rationales enter the target-level
adjudication but not the deterministic path-coverage score.

Four cases require more reference files than the six-target cap permits.
They remain in every all-case measure, in \Cref{tab:action52-main} and in \Cref{tab:action52-complete}.
For completeness, Table~\ref{tab:action52-complete} also gives all-file
success on the 48 cap-eligible cases. The stored field calls this \texttt{strict\_file\_success}, which denotes a different denominator rather than full patch-test success.

%% No float page for this one: with \@fpbot stretching at the foot, a half-full float
%% page wastes the remainder in a single block. Measured at 317pt on page 26.
\begin{table}[htb]
\centering
\small
\caption{Additional Action52 accounting. Complete count is out of all 52 cases;
cap-eligible success uses the 48 cases whose reference file set fits the
target cap. Reference-file recall is the task-macro mean. All ten conditions use
the same states.}
\label{tab:action52-complete}
\setlength{\tabcolsep}{6pt}
\begin{tabular*}{\textwidth}{@{\extracolsep{\fill}}llrrr@{}}
\toprule
Executor & Evidence & \shortstack{Complete\\count} &
\shortstack{Cap-eligible\\success} & \shortstack{Reference-file\\recall (\%)} \\
\midrule
DeepSeek & Oracle $-$ 1 group & 0/52 & 32/48 & 68.53 \\
DeepSeek & Qwen3 Embedding & 11/52 & 33/48 & 70.32 \\
DeepSeek & Qwen3 + Reranker & 17/52 & 34/48 & 71.56 \\
DeepSeek & \msscomplement & 27/52 & 36/48 & 76.01 \\
DeepSeek & Oracle-MSS & 52/52 & 38/48 & 80.90 \\
\midrule
Claude & Oracle $-$ 1 group & 0/52 & 37/48 & 78.84 \\
Claude & Qwen3 Embedding & 11/52 & 36/48 & 77.96 \\
Claude & Qwen3 + Reranker & 17/52 & 36/48 & 77.35 \\
Claude & \msscomplement & 27/52 & 38/48 & 79.60 \\
Claude & Oracle-MSS & 52/52 & 39/48 & 82.34 \\
\bottomrule
\end{tabular*}
\end{table}

%% No float page for this one: with \@fpbot stretching at the foot, a half-full float
%% page wastes the remainder in a single block. Measured at 317pt on page 26.
\begin{table*}[htb]
\centering
\small
\caption{Action52 target-level adjudication. Raw precision uses the number of
targets actually emitted. Fixed-budget target precision uses the common 171-slot
denominator and is the equal-budget value reported in
\Cref{tab:action52-main}. ``Other'' denotes an explicit non-precise verdict
without a forced coarse-versus-wrong subtype.}
\label{tab:action52-target-detail}
\setlength{\tabcolsep}{3.5pt}
\begin{tabular*}{\textwidth}{@{\extracolsep{\fill}}llrrrrr@{}}
\toprule
Executor & Evidence & Precise & Coarse & Wrong & Other &
Raw precision (\%) \\
\midrule
DeepSeek & Qwen3 Embedding & 83 & 62 & 22 & 0 & 49.70 \\
DeepSeek & Qwen3 + Reranker & 88 & 57 & 17 & 0 & 54.32 \\
DeepSeek & Oracle $-$ 1 group & 90 & 55 & 18 & 0 & 55.21 \\
DeepSeek & \msscomplement & 110 & 52 & 7 & 0 & 65.09 \\
DeepSeek & Oracle-MSS & 111 & 54 & 2 & 0 & 66.47 \\
\midrule
Claude & Qwen3 Embedding & 93 & 61 & 8 & 0 & 57.41 \\
Claude & Qwen3 + Reranker & 93 & 56 & 12 & 0 & 57.76 \\
Claude & Oracle $-$ 1 group & 95 & 55 & 10 & 4 & 57.93 \\
Claude & \msscomplement & 100 & 54 & 6 & 2 & 61.73 \\
Claude & Oracle-MSS & 114 & 42 & 5 & 0 & 70.81 \\
\bottomrule
\end{tabular*}
\end{table*}

\paragraph{Paired effects and interpretation.}
Within each executor family, the state, decision prompt, cap, and scorer are
fixed while the evidence changes. Across the retrieval and Oracle-MSS rows, the
fixed-budget target-precision estimates follow the listed group-recall ordering
in both families. The missing-group control is the exception: its group recall
is below the reranked baseline's while its precision is above it under both
executors. The Qwen, \msscomplement, and Oracle rows also increase
in Complete-MSS, mean reference-file recall, and all-file success.
DeepSeek's
full-coverage sequence over all 52 cases is 63.46\%, 69.23\%, and 73.08\% at completeness
levels 21.15\%, 51.92\%, and 100\%; Claude's is 69.23\%, 73.08\%, and
75.00\% at the same levels. These percentages use all 52 cases, including the four cases exceeding the six-target cap. The cap-eligible success column instead uses 48 cases with the same numerators.

The Oracle-minus-one condition supplies a paired evidence-removal check. Its
Complete-MSS is zero because every task excludes one group's registered
alternatives, although it retains partial evidence and 48.27\% group recall.
Against the same-count Oracle, removing that one group costs 12.28 points of fixed-budget precision with DeepSeek and 11.11 with Claude, and lowers mean reference-file recall for both executor families. The executor still reads the visible state and interprets the target locations, so these retrieval metrics describe the evidence supplied to localization while target decisions remain executor outputs.

Table~\ref{tab:action52-paired} retains uncertainty at the task level.
We resample the same 52 paired task identifiers 10,000 times with seed
20260904 and report percentile intervals for mean reference-file recall.
Each repository contributes one task. The DeepSeek oracle-group contrast has a wholly positive interval. Both learned-retrieval contrasts have positive point estimates with intervals spanning zero at this cohort size. \Cref{tab:action52-reranker-paired} applies the same procedure to the reranked control and adds fixed-slot target precision. Pairing against the reranked condition is tighter than against plain embedding because the two selections agree on most cases, so the \msscomplement contrast separates there on a smaller point estimate.

\begin{table}[htb]
\centering
\small
\caption{Paired Action52 differences in mean reference-file recall.
Intervals are in percentage points and resample tasks, not condition means.}
\label{tab:action52-paired}
\begin{tabular*}{\textwidth}{@{\extracolsep{\fill}}llrr@{}}
\toprule
Executor & Evidence comparison & Difference (pp) & 95\% interval \\
\midrule
DeepSeek & MSS $-$ Qwen & +5.69 & $[-2.41,+14.26]$ \\
DeepSeek & Oracle $-$ MSS & +4.89 & $[-1.92,+12.58]$ \\
DeepSeek & Oracle $-$ missing group & +12.36 & $[+4.18,+21.81]$ \\
Claude & MSS $-$ Qwen & +1.64 & $[-3.53,+8.05]$ \\
Claude & Oracle $-$ MSS & +2.74 & $[+0.00,+7.40]$ \\
Claude & Oracle $-$ missing group & +3.50 & $[-1.44,+9.76]$ \\
\bottomrule
\end{tabular*}
\end{table}

\begin{table}[htb]
\centering
\small
\caption{Paired Action52 differences against the reranked control, matching
\Cref{tab:action52-paired}; the target-precision block also gives Oracle $-$ missing group. Differences are in
percentage points. W/T/L counts cases on which the left condition is better, equal, or worse.}
\label{tab:action52-reranker-paired}
\setlength{\tabcolsep}{4pt}
\begin{tabular*}{\textwidth}{@{\extracolsep{\fill}}llrrr@{}}
\toprule
Executor & Comparison & Difference (pp) & 95\% interval & W/T/L \\
\midrule
\multicolumn{5}{l}{\emph{Reference-file recall}} \\
DeepSeek & Reranker $-$ Embedding & $+1.24$ & $[-5.17,+7.69]$ & 2/47/3 \\
DeepSeek & \msscomplement $-$ Reranker & $+4.44$ & $[+0.12,+10.44]$ & 4/48/0 \\
Claude & Reranker $-$ Embedding & $-0.60$ & $[-6.33,+5.17]$ & 2/48/2 \\
Claude & \msscomplement $-$ Reranker & $+2.24$ & $[-2.88,+8.65]$ & 2/48/2 \\
\addlinespace[2pt]
\multicolumn{5}{l}{\emph{Fixed-slot target precision}} \\
DeepSeek & Reranker $-$ Embedding & $+2.92$ & $[-5.14,+11.45]$ & 14/27/11 \\
DeepSeek & \msscomplement $-$ Reranker & $+12.87$ & $[+5.39,+20.36]$ & 18/29/5 \\
DeepSeek & Oracle $-$ missing group & $+12.28$ & $[+4.27,+20.34]$ & 20/26/6 \\
Claude & Reranker $-$ Embedding & $0.00$ & $[-6.83,+6.63]$ & 11/30/11 \\
Claude & \msscomplement $-$ Reranker & $+4.09$ & $[-1.89,+10.29]$ & 13/32/7 \\
Claude & Oracle $-$ missing group & $+11.11$ & $[+5.20,+17.51]$ & 15/35/2 \\
\bottomrule
\end{tabular*}
\end{table}

\FloatBarrier
\subsection{End-to-End Repair on Fresh23}
\label{app:fresh23}

This appendix gives the full accounting behind \Cref{tab:fresh23-body}. Fresh23 is a frozen
23-task cohort of 21 Windows tasks and two macOS tasks, one task per repository. Each task runs the whole loop. The retrieval method
supplies evidence, the executor proposes a patch, the patch is applied to an executable
repository snapshot, and the repository's own tests decide the outcome. Every condition covers
all 23 tasks, issues two or three iterative retrieval calls per task, returns at most five
source units per call, and has no retrieval fallback. The per-task executor assignment is
identical across conditions: 20 tasks use DeepSeek-v4-pro and three use
DeepSeek-v4-flash. Strict resolution requires the fail-to-pass (F2P) tests to pass and the selected
pass-to-pass (P2P) neighborhood to stay green; pass-to-pass is evaluated only where fail-to-pass
passes.

The three retrieval conditions read no organizer-private label at inference time. The two
Oracle conditions are organizer-constructed diagnostics that do read the private certificate,
and they are reported as bounds on the target rather than as deployable methods. Oracle-MSS
covers every required group; the minus-one-group condition removes one required group's
registered alternatives from that same evidence and refills the vacated slots, so evidence
volume and the rest of the pipeline are held fixed while certificate completeness is broken.

\begin{table}[htbp]
\centering
\caption{Fresh23 end-to-end repair over 23 frozen tasks. Resolved columns are counts with
percentages. F2P is fail-to-pass and P2P pass-to-pass, and P2P is evaluated only where F2P passes. Candidate target-file hit counts the tasks in which some attempted patch modifies at least one file the reference repair touches, including attempts that were not ultimately applied. Tokens are the executor's mean per-task total, prompt plus completion, and calls are the mean
iterative retrieval calls. Bold marks the best resolution count among the three deployable conditions.}
\label{tab:fresh23-e2e}
\small
\setlength{\tabcolsep}{4pt}
\begin{tabular*}{\textwidth}{@{\extracolsep{\fill}}lrrrrrr@{}}
\toprule
& \multicolumn{3}{c}{Resolved of 23} & & \multicolumn{2}{c}{Cost} \\
\cmidrule(lr){2-4}\cmidrule(lr){6-7}
Condition & Strict & F2P & P2P & \shortstack{Candidate\\target file} & Tokens & Calls \\
\midrule
\multicolumn{7}{l}{\emph{Retrieval, no private label at inference time}} \\
BM25 & 5 (21.74) & 8 (34.78) & 6 & 22 & 133,775 & 2.52 \\
Qwen3 + Reranker & 6 (26.09) & \best{10 (43.48)} & 7 & 20 & 126,261 & 2.43 \\
\msscomplement & \best{8 (34.78)} & \best{10 (43.48)} & \best{9} & 20 & 118,351 & 2.52 \\
\addlinespace[2pt]
\multicolumn{7}{l}{\emph{Organizer-private certificate diagnostics}} \\
Oracle $-$ 1 group & 7 (30.43) & 8 (34.78) & 8 & 23 & 98,121 & 2.48 \\
Oracle-MSS & 8 (34.78) & 11 (47.83) & 9 & 23 & 99,655 & 2.48 \\
\bottomrule
\end{tabular*}
\end{table}

\begin{table}[htbp]
\centering
\caption{Paired Fresh23 contrasts on the same 23 tasks, with 20,000 task-clustered bootstrap
draws per row and seed 20260912. Win, tie and loss count the tasks on which the first condition
resolves and the second does not, both agree, and the second resolves and the first does not.}
\label{tab:fresh23-paired}
\small
\setlength{\tabcolsep}{6pt}
\begin{tabular*}{\textwidth}{@{\extracolsep{\fill}}lrrr@{}}
\toprule
Contrast & $\Delta$ (points) & 95\% interval & W/T/L \\
\midrule
\multicolumn{4}{l}{\emph{Strict resolution}} \\
\msscomplement vs Qwen3 + Reranker & $+8.70$ & $[-13.04,+30.43]$ & 4/17/2 \\
\msscomplement vs BM25 & $+13.04$ & $[-4.35,+30.43]$ & 4/18/1 \\
Oracle-MSS vs Oracle $-$ 1 group & $+4.35$ & $[-8.70,+17.39]$ & 2/20/1 \\
Qwen3 + Reranker vs BM25 & $+4.35$ & $[+0.00,+13.04]$ & 1/22/0 \\
\addlinespace[2pt]
\multicolumn{4}{l}{\emph{Fail-to-pass resolution}} \\
Oracle-MSS vs Oracle $-$ 1 group & $+13.04$ & $[+0.00,+26.09]$ & 3/20/0 \\
Oracle-MSS vs BM25 & $+13.04$ & $[+0.00,+26.09]$ & 3/20/0 \\
\msscomplement vs BM25 & $+8.70$ & $[-8.70,+26.09]$ & 3/19/1 \\
\msscomplement vs Qwen3 + Reranker & $+0.00$ & $[-17.39,+17.39]$ & 2/19/2 \\
\bottomrule
\end{tabular*}
\end{table}

The minus-one-group contrast is the cleanest reading of the target on this cohort, and \Cref{tab:fresh23-paired} gives every paired contrast with its interval. The two
Oracle conditions share an executor, a token ceiling and an evidence volume, and differ only in
whether one required group is covered. Removing it costs three fail-to-pass resolutions and
returns the condition to BM25's eight, and under the fail-to-pass criterion no task is resolved by the incomplete certificate but missed by the complete one.

\msscomplement matches Oracle-MSS on strict resolution. Its fail-to-pass count ties the reranked baseline at ten, but its pass-to-pass count is nine against seven, so its repairs keep the surrounding tests passing more often, which is what the strict criterion records. The cohort tests directly whether the benchmark's target predicts an outcome it does not control.

Candidate target-file hit does not order the conditions the way resolution does. BM25 reaches 22 of 23 while resolving five, \msscomplement and the reranked baseline both reach 20, and the Oracle conditions reach 23. Attempting an edit in a file the repair touches is not the same as covering what the decision requires, which is what the certificate records and what the resolution columns follow. One task, \texttt{graphistry\_\_pygraphistry-661}, has an incomplete pass-to-pass environment and is therefore unresolved under the strict criterion in every condition even though its fail-to-pass and pass-to-pass checks pass throughout; that single task is why the pass-to-pass count in \Cref{tab:fresh23-e2e} exceeds the strict count by exactly one in every row. Wall-clock latency is
recorded in the released runs but is not compared here: the reranked baseline serves its
reranker locally while the remaining components call hosted endpoints, so its durations
describe that deployment rather than the method. The token counts in \Cref{tab:fresh23-e2e} are
deployment independent and carry the cost comparison.

\FloatBarrier
\section{Artifact, Responsible Use, and Reproduction}
\label{app:artifacts}

The release separates the supplementary research archive from an organizer-private evaluator that is not distributed with it. The supplementary archive contains 500 labeled Cal500 states and 500 label-free Test500 states, together with sanitized candidate pools, schemas, scoring code, and reconstruction metadata; sanitization removes private fields, identities, and secrets, and does not filter observed evidence. The archive also carries the state-level observed-overlap assignments, the twelve-method stratified matrix, the aggregation script, and its source hashes and seed. Test500 public files contain no grouped certificate, acceptable alias set, necessity weight, gold role, next-action label, organizer-private filename, API key, gold patch, or Python bytecode. The organizer archive contains exactly 500 Test500 certificates together with reserve and excluded review records, and is marked not for publication.

Clean-environment reproduction covers the public scorer, SSR baseline, and a three-state \msscomplement smoke run. Full semantic evaluation requires a compatible structured-output endpoint. Candidate generation and scoring run locally in the declared Python/PyTorch environment. Frozen result JSON/CSV files, configuration contracts, and the separate reproducibility ledger map every reported result to its source artifact.

The Fresh23 end-to-end cohort ships as a separate frozen package. It holds one run record per method and task carrying the proposed patch, the executed test commands and their outcomes, and the retrieval trace behind the supplied evidence, together with a per-task comparison table, an aggregate summary, and a SHA-256 digest for every file. Repository snapshots are identified by commit for reconstruction under upstream terms rather than bundled. It also carries the run records of the two organizer-constructed certificate diagnostics; the certificates those diagnostics read are not distributed. No deployable method reads an organizer-private label at inference time.

\section{Scope, Limitations, and Release Responsibilities}
\label{app:scope}

SERBench focuses on repository evidence from multi-language coding trajectories. The visible state schema and grouped-MSS scorer apply across agent frameworks and programming languages. All test states, however, come from one trajectory generator, so other generators and states without the need, hypothesis, and subgoal fields remain untested. Certificates specify sufficient alternatives for the current decision, allowing multiple minimal solutions and leaving the agent's internal reasoning unconstrained. This target complements long-context models: a larger memory remains available for discovery, while the selected working set is evaluated against the decision's unresolved requirements. The experiments measure this selection and its downstream use.

Model-assisted annotation retains item-level provenance and independent verification. Repository source remains governed by upstream licenses, and the release distributes reconstruction information rather than bundled snapshots, as described in \Cref{app:artifacts}.

The paired action cohorts keep states, repository snapshots, tools, prompts, and evaluators fixed across evidence conditions. They connect state-conditioned evidence recovery to an application endpoint without changing the benchmark target. Private certificates remain held out from training.

\end{document}